\newcommand*{\chni}{}
\newcommand*{\chn}{}
\newcommand*{\cht}{}
\newcommand*{\chel}{}

\newcommand*{\chtw}{}
\newcommand*{\chth}{}
\newcommand*{\chft}{}

\documentclass[twocolumn]{aastex62}

\usepackage{savesym}
\usepackage{amsmath,amstext}
\usepackage{gensymb}
\let\splitbox\relax 
\usepackage{blindtext}
\usepackage{rotating}
\usepackage[caption=false]{subfig}
\usepackage{CJK}

\def\mJybeam{mJy\,beam$^{-1}$}

\def\as{\farcs}

\def\C18O{C$^{18}$O}
\def\3CO{$^{13}$CO}
\def\2CO{$^{12}$CO}

\def\N2Dp{N$_2$D$^+$}

\graphicspath{{./}{figures/}}

\submitjournal{ApJ}

\shorttitle{Rotating filament in Orion B}
\shortauthors{Hsieh et al.}

\begin{document}
\begin{CJK*}{UTF8}{gbsn}

\title{
Rotating 
filament in
Orion B: 
Do cores inherit their angular momentum from their parent filament?}

\correspondingauthor{Cheng-Han Hsieh}
\email{cheng-han.hsieh@yale.edu}

\author[0000-0003-2803-6358]{Cheng-Han Hsieh (承翰)}
\affiliation{Department of Astronomy, Yale University, New Haven, CT 06511, USA}


\author{H{\'e}ctor G. Arce}
\affiliation{Department of Astronomy, Yale University, New Haven, CT 06511, USA}

\author{Diego Mardones}
\affiliation{Departamento de Astronomia, Universidad de Chile, Casilla 36-D, Santiago, Chile}

\author{Shuo Kong}
\affiliation{Department of Astronomy, Yale University, New Haven, CT 06511, USA}
\affiliation{Steward Observatory, University of Arizona, Tucson, AZ 85719, USA }

\author{Adele Plunkett}
\affiliation{National Radio Astronomy Observatory, 520 Edgemont Rd, Charlottesville, VA 22903, USA}




\begin{abstract}

Angular momentum is one of the most important physical quantities that governs star formation. The initial angular momentum of a core may be  responsible for its fragmentation and can have an influence on the size of the protoplanetary disk. To understand how cores obtain their initial angular momentum, it is important to study the angular momentum of filaments where they form. While theoretical studies on filament rotation have been explored,  there exist very few observational measurements of the specific angular momentum in star forming filaments. 
{\chtw We} present high resolution \N2Dp ALMA observations 
of the LBS\,23 (HH24-HH26) region in Orion B, which  provide one of the most reliable measurements of the specific angular momentum in a star-forming filament. We find the total specific angular momentum ($4 \times 10^{20}$\, cm$^{2}$\,s$^{-1}$),
the dependence of the specific angular momentum with radius ({\chtw $j(r) \propto r^{1.83}$), and the ratio of rotational energy to gravitational energy ($\beta_{rot} \sim 0.04$) }
comparable to those observed in
rotating cores with sizes similar to our filament width ($\sim$0.04\,pc) in other star-forming regions.  {\chtw Our filament angular momentum profile is consistent with rotation acquired from ambient turbulence and 
with simulations that show cores and their host filaments  develop simultaneously due to multi-scale growth of nonlinear perturbation generated by turbulence.
} 




\end{abstract}


\section{Introduction} 
\label{sec:intro}
\end{CJK*}


One of the main challenges in star formation is trying to solve the so-called ``angular momentum problem",  which arises from the fact that the observed angular momentum of individual stars is much smaller than that of  molecular cloud cores from which they presumably formed \citep{2003RPPh...66.1651L,2004A&A...423....1J}. At large {\chel molecular cloud scales ($\sim 10$\,pc)}, where the density is low and the  ionization fraction is high {\chel ($\sim 10^{-6}$) \citep{2005pcim.book.....T}}, {\chni magnetic braking effects may be effective in removing most of the angular momentum} \citep{1994ApJ...432..720B,1999ApJ...520..706C}. At smaller scales, {\chni  studies conducted about two to three decades ago
found the specific angular momentum\footnote{\chtw Specific angular momentum is defined as angular momentum per unit mass.} of cores is comparable to that of wide separation binaries ($\sim$\,1000\,AU)
\citep{1993ApJ...406..528G,1998ApJ...504..207B,1999ApJS..125..161J,2002ApJ...572..238C}.  Other observations  also found that at scales smaller than 0.03\,pc the angular momentum is conserved \citep{1997ApJ...488..317O,2000prpl.conf..217M,2002A&A...393..927B,2013EAS....62...25B,2014prpl.conf..173L}, which is consistent with the theoretical  prediction of weaker magnetic braking effects at smaller scales \citep{1997ApJ...485..240B}. {\chni The question of how cores obtain their initial angular momentum remains  an important and heated topic today in star formation as it sets the angular momentum budget for the formation of multiple systems and protoplanetary disks.}


{\chni Recent Hershel studies have shown that most stars form in filaments with a supposed universal width of 0.1\,pc \citep{2010A&A...518L.102A,2011A&A...529L...6A,2019A&A...621A..42A,2015MNRAS.452.3435K,2015A&A...574A.104T}.  If most  cores are formed in filaments {\chel \citep{2014prpl.conf...27A}}, one would expect then that filament rotation and fragmentation could play an important role in explaining the origin of core angular momentum.}



{\chni One possible explanation for the origin of core angular momentum is turbulence. {\chn A recent theoretical study by} \citet{2019ApJ...881...11M} found the observed dependence of specific angular momentum with mass can be explained by one-dimensional Kolmogorov {\chn (isotropic)} turbulence perturbations in filaments. While their results are consistent with observations, this model is likely far from complete as it does not incorporate magnetic fields which are thought to be important at filament scales {\chn \citep{2013A&A...550A..38P}} {\chn and result in anisotropic turbulence.} 
Moreover, \citet{2019ApJ...881...11M} does not consider how  the global filament rotation affects the initial angular momentum of cores. }

{\chni While a number of numerical studies show that turbulence may provide the initial angular momentum in cores \citep[e.g.,][]{2000ApJ...543..822B,2018ApJ...865...34C,2019ApJ...881...11M}, {\chtw some} simulations underestimate the core angular momentum by a factor of 10 compared to values derived from observations {\chel \citep[e.g.,][]{2008ApJ...686.1174O,2010ApJ...723..425D}}. {\chn \citet{2010ApJ...723..425D} compared the intrinsic angular momentum ($j_{3D}$) in their simulated cores to synthetic specific angular momentum derived from  2-dimensional velocity maps ($j_{2D}$) and found $j_{3D}$/$j_{2D}$ $\sim0.1$, which suggest observations overestimate the true angular momenta by an order of magnitude}.  However, \citet{2018MNRAS.480.5495Z} suggest this order of magnitude difference does not come from observational error but possibly due to numerical effects. More recently,  \citet{2019ApJ...876...33K} conducted a conservative order of magnitude calculation
and showed the resulting angular momentum is an order of magnitude below the minimum angular momentum observed in cores, {\chn possibly} indicating that the model of 
turbulent origin for the angular momentum in cores is inconsistent with observations. {\chn Yet, \citet{2019ApJ...881...11M}'s theoretical turbulent models tend to reproduce observations well with reasonable parameters. A turbulence-induced origin for the angular momentum in cores cannot yet be ruled out.}

{\chni In contrast with most previous studies, \citet{2019ApJ...876...33K} propose that the initial angular momentum of cores is generated locally and comes from the gravitational interactions of overdensities (dense cores), and it is  not inherited from the large scale initial cloud rotation. Thus there is still no consensus on the 
origin of angular momentum in cores. 

Filaments, which lie at intermediate scales between molecular cloud scales ($\sim$10\,pc) and the small core/envelope scales ($\sim$0.01\,pc) 
might be the key for understanding the origin of a core's initial angular momentum.}}
Observations of the L1251 infrared dark cloud found that this
$\sim 3.3$\, pc $\times0.3$\,pc 
filament is rotating along its minor axis with an angular frequency ($\omega$) of about $7 \times 10^{-15}$\,rad\,s$^{-1}$  \citep{2016A&A...586A.126L}. 
{\chtw In addition to L1251, only a few more filaments have been observed to show a velocity gradient along the filament minor axis, but in most cases the gradients have  been interpreted as being caused by accretion in a flattened structure, 
converging accretion flows or multiple components aligned in the line of sight, and not by rotation
\citep{2014ApJ...790L..19F,2015A&A...584A..67B,2018ApJ...853..169D,2020MNRAS.494.3675C}.} Due to the difficulty in identifying filaments in line maps, the requirement of large high-sensitivity maps of an optically thin line with both high velocity and angular resolution, and the possible degeneracy in interpreting the complicated motion within a filament, {\cht precise measurements of the rotation and  angular momentum of filaments are very rare.} 

{\cht In this {\chtw paper}, we present 
detailed measurements of the specific angular momentum of a star-forming filament. The area we studied, the HH24-26 {\chel  low-mass star forming region} (a.k.a., LBS23) is a 1\,pc long filament located in Orion B  with a total mass around $230$\,$M_\odot$ and at a distance of approximately $400\,$pc \citep{1991ApJ...368..432L,1999ApJ...527..856L}. 
The large fraction of Class 0 protostars ($\sim $50\,\%) compared to the more evolved Class I, II and III sources in this region \citep[see][]{2012AJ....144..192M,2016ApJS..224....5F}} suggest that LBS23 is a very young  filament undergoing its first major phase of star formation.


In the following section {\chtw (Section \ref{sec:obs})} we describe the observational data, the calibration, and the imaging process. In Section \ref{sec:results} we show the results of our observation. In Section \ref{sec:discussion}, we analyse and discuss the stability of the {\chni rotating} filament, its dynamics and physical properties (density profile, turbulence, {\chtw magnetic fields, specific angular momentum and energy}). In Section \ref{sec:conclusion} we summarize the main findings and give our conclusions. 

\section{Observations} 
\label{sec:obs}

The results presented here come from Cycle 4  Atacama Large Millimeter/submillimeter Array (ALMA) observations of the LBS23 region
(project ID: {\chel 2016.1.01338.S}, PI: D.~Mardones). {\chtw The original goal of the project was to characterize the kinematics of gas and stellar feedback, over a range of scales, in a young filamentary cloud with active star formation. The observations were conducted using one spectral configuration (using ALMA Band 6) which simultaneously observed the 1.29 mm dust continuum emission and the following six molecular lines which trace different density and kinematic regimes:  
$^{12}$CO(2-1), $^{13}$CO(2-1), C$^{18}$O(2-1), H$_2$CO($3_{0,3}$-$2_{0,2}$), SiO(5-4), and \N2Dp (3-2). 
In this paper we concentrate on the 
\N2Dp  (3-2) line (with a rest frequency of 231.32 GHz), the highest-density tracer in our spectral set-up as it clearly shows the most complete structure of the high-density narrow filament in the LBS23 region. Deuterated molecules such as \N2Dp trace cold and dense gas and is a good tracer of dense structures like cores and filaments \citep{2010ApJ...718..666F,2015ApJ...804...98K,2017ApJ...834..193K}.
Other lines which trace the more diffuse ambient gas, outflows and their impact on the cloud will be presented in future papers.}

We used two mosaic fields to cover the area of interest. The northern field is {\chn 220\arcsec\, by 110\arcsec} and it is centered at 05\textsuperscript{h}46\textsuperscript{m}08\fs16, -00\degree10\arcmin58\as80 (J2000). The southern field is {\chn 150\arcsec \, by 110\arcsec} and is centered  at 
05\textsuperscript{h}46\textsuperscript{m}06\fs00, -00\degree13\arcmin49\as80 (J2000).
In total these two fields encompass an area of about {\chn 360\arcsec \,by\, 110\arcsec}, which was observed using the ALMA 12 m, 7 m and Total Power (TP) arrays.


To cover the northern field, we used 116 pointings with the 12\,m array in the C34-1 configuration  and 42 pointings with the 7\,m array {\cht between} 2017 March 26 and {\cht 2017 April 28}. We used J0750+1231 as the calibrator for bandpass and flux calibration and J0552+0313 for phase calibration of  the 12\,m array visibility data. The total on-source integration time of the 12\,m array, made with two executions, was 69 minutes and sampled baseline ranging between approximately 15\,m to 390\,m. The 7\,m array observations, which were made with {\chtw 10} executions with a total on-source integration of 141\,minutes, used J0522-3627 for bandpass and flux calibration and J0532+0732 for phase calibration. The {\chn baseline coverage was from  about 7\,m\, to 49\,m.}

For the southern field, we observed 84 pointings with the 12\,m array and 32 pointings with the 7\,m array {\cht between} 2017 March 22 and 2017 {\cht April 16}. For the 12\,m array observation, the baseline ranged from about 15\,m to  161\,m and two executions provided a  total on-source time of around 54 minutes. As for the 7\,m array observations, the total on-source time was 70 minutes (provided by 7 different executions), and employed  baselines ranging between about 9\,m to  49\,m. The southern field observations used the same calibrators as the northern field observations mentioned above. For both fields we utilized the ALMA pipeline in the Common Astronomy Software Applications (CASA) version 5.4.0-70 to  calibrate the data.



The northern and southern regions were observed with the total power array between late March and early July of 2017.
The northern region was mapped with 2 to 4 antennas operated as single dish telescopes with 26 executions and the total on-source time was about 25 hrs. 
As for the southern region, it was mapped with 3 to 4 TP antennas with 16 executions and the total on-source time was about 17 hrs.
The TP array data exhibit a deep telluric absorption feature near the \N2Dp line, which hampers the ability of the CASA pipeline to do a proper baseline subtraction. We therefore conducted our own baseline subtraction by fitting the baseline within $\pm$\,20 km s$^{-1}$ of the  \N2Dp line and at frequencies {\chn higher} than the atmospheric absorption peak with a second degree polynomial. 



To recover the extended emission as much as possible, we used the properly baseline-subtracted total power map as both a start model and a mask
when imaging the interferometry data using the CASA task \textit{tclean}. We used the multi-scale deconvoler with scales of 0,\,1,\,3 beam sizes  to recover flux at different scales, and natural weighting was applied to achieve the highest signal to noise ratio possible. To obtain a uniform synthesized beam across the entire area covered, we restricted the UV data to include baselines of  up to 158\,m for both fields. We used the CASA task  \textit{FEATHER} to combine the 12\,m and 7\,m array (interferometer) data with the total power data. The resulting  map has a  beam of 2\as66 $\times$ 1\as61 (P.A. = -64.74\degree) with a rms noise level of 32 \mJybeam \/ {\chn per channel} and  a velocity resolution of 0.08 km\,s$^{-1}$.

The 1.29\,mm continuum dust emission was observed (with bandwidth of 1875\,MHz)
simultaneously with the \N2Dp line and calibrated using the same pipeline. Interactive cleaning with natural weighting was used for imaging the combined 12m and 7m array data, resulting in a dust continuum map with a synthesized beam of 2\as64 $\times$ 1\as58 (P.A. = -64.80\degree) and a rms noise level of 0.54 \mJybeam.


\section{Results} 
\label{sec:results}

\begin{figure*}[tbh]
\centering
\makebox[\textwidth]{\includegraphics[width=\textwidth]{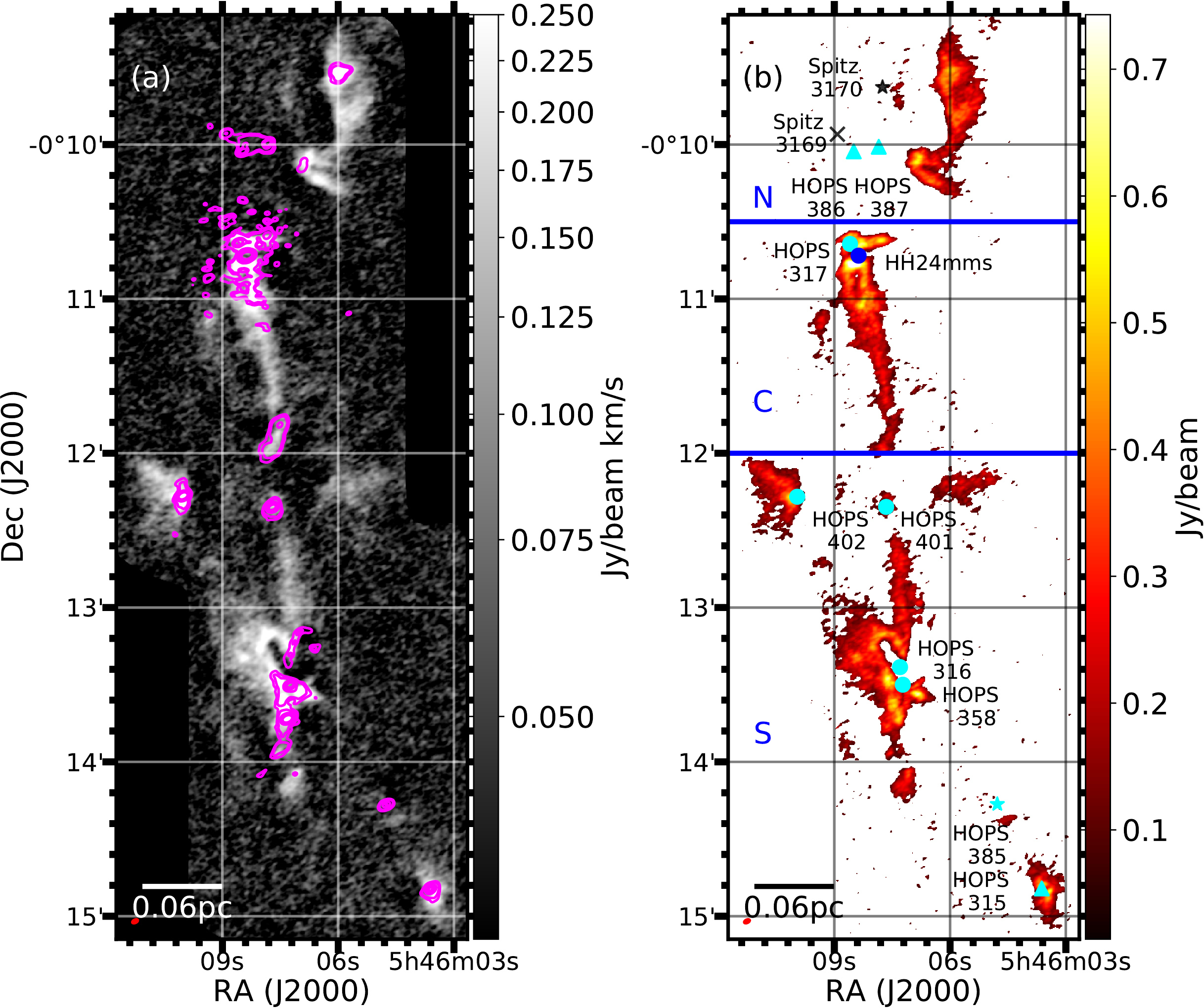}}
\caption{ ALMA \N2Dp $J = 3-2$ maps of the LBS23 region.
(a) 
Integrated intensity map of the \N2Dp emission. The magenta contours represent the 1.29\,mm dust continuum emission in steps of 3$\sigma$, 5$\sigma$, 20$\sigma$, 40$\sigma$, 80$\sigma$, 320$\sigma$, where $\sigma = 5.4\times 10^{-4}$ Jy beam$^{-1}$. (b) \N2Dp peak intensity  (moment 8) map. Previously known sources are shown. The HOPS Class 0, Class I and flat-spectrum sources are marked by cyan circles, cyan triangles and cyan stars, respectively {\chel \citep{2016ApJS..224....5F}}. {\cht The Spitzer-identified pre-main sequence star and red protostar (detection in 24$\mu$m without detection in  4.5, 5.8 or 8$\mu$m)
are marked as a black star and cross, respectively {\chel \citep{2012AJ....144..192M}}. {\chtw The blue circle marks the position of the Class 0 protostar, HH24\,mms.} The northern, central and southern regions are identified with the letters N, C, and S, respectively, and  separated by thick blue horizontal lines. 
The size of the synthesized beam is represented with a red ellipse in the lower left corner of each panel.}
}
\label{fig1}
\end{figure*}

\begin{figure*}[tbh]
\centering
\makebox[\textwidth]{\includegraphics[width=\textwidth]{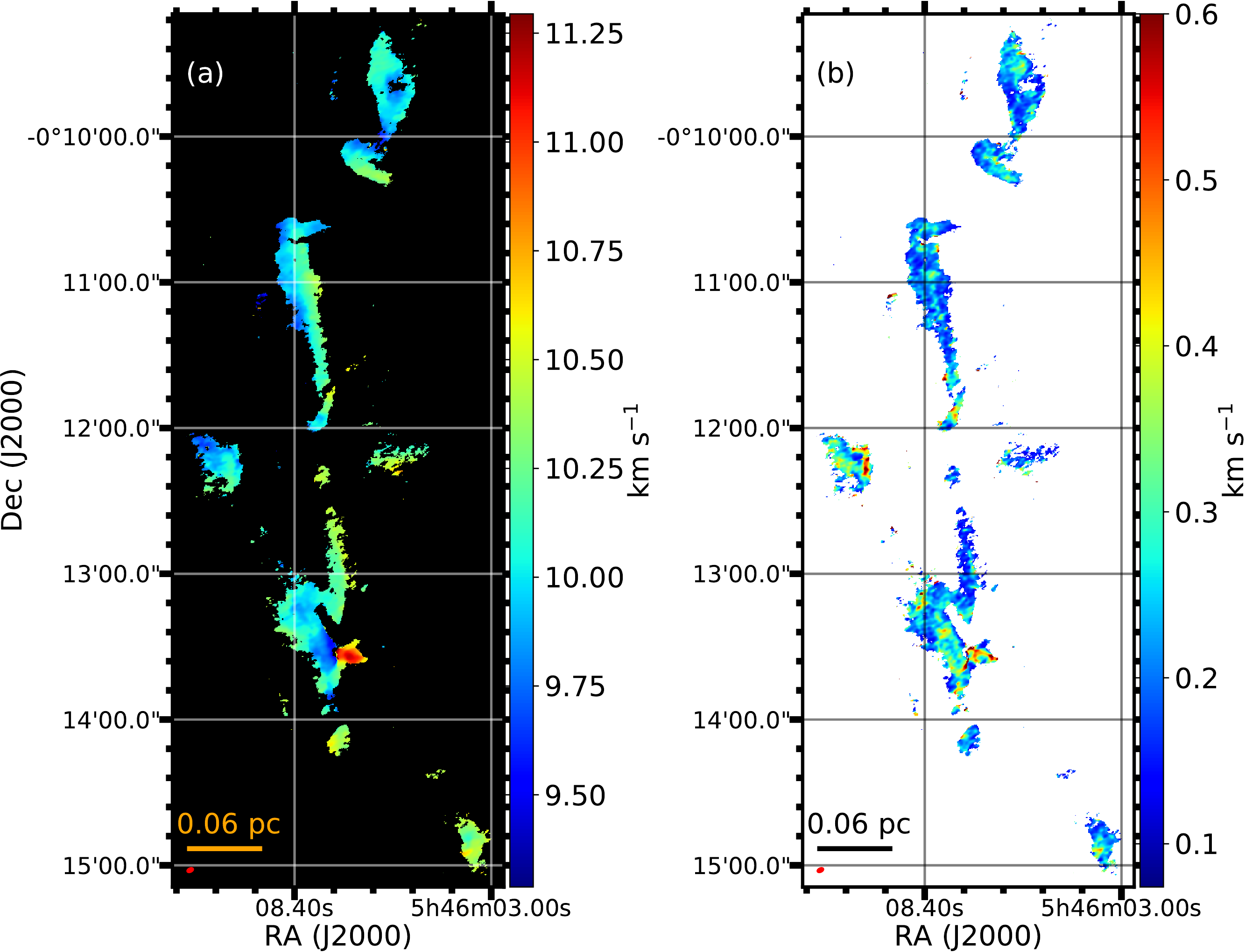}}
\caption{ Kinematics of the entire LBS23 region from fits to the \N2Dp  hyperfine line emission.
(a) System velocity map. (b) Velocity dispersion map.
The  beam is shown as a filled red ellipse in the lower left corner of each panel.
}
\label{fig2}
\end{figure*}

\begin{figure*}[tbh]
\centering
\makebox[\textwidth]{\includegraphics[width=\textwidth]{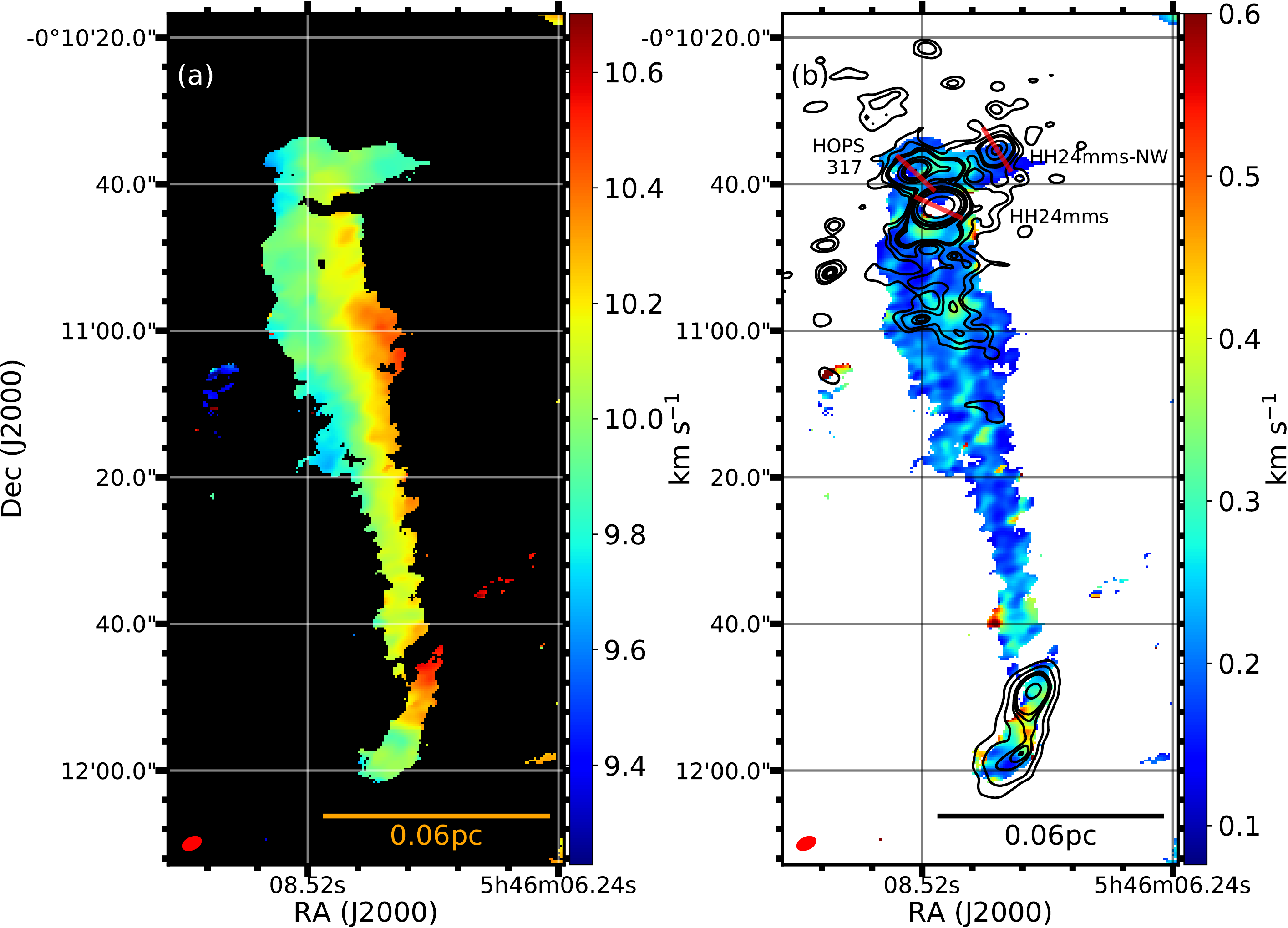}}
\caption{Kinematics in the central filament in LBS23. These panels are zoomed-in versions of the panels in Figure 2 (with a different color scale). 
(a) System velocity map. (b). Velocity dispersion map. The black contours in panel b show the 1.29\,mm dust continuum emission in steps of 3$\sigma$, 5$\sigma$, 7$\sigma$, 9$\sigma$, 10$\sigma$, 20$\sigma$, 40$\sigma$, 80$\sigma$, 320$\sigma$, where $\sigma = 5.4\times 10^{-4}$ Jy beam$^{-1}$. The red lines mark the direction of the outflows powered by the three protostars identified in the northern edge of the filament. 
The size of the synthesized beam is represented with a red ellipse in the lower left corner of each panel.
 }
\label{fig3}
\end{figure*}

The \N2Dp integrated intensity map is shown in  \autoref{fig1}a. Most of the emission is concentrated in filamentary structures elongated along the north-south direction. We identify three different regions:
the north, central and south regions. In \autoref{fig1}b these are separated by thick horizontal {\chel blue} lines and denoted as N, C and S. The previously known young stellar objects 
in this area are also shown in \autoref{fig1}b. {\chtw The maps of the system velocity and velocity dispersion are shown in \autoref{fig2}.}


The northern region includes a {\chn 65\arcsec (0.13\,pc)}-long filamentary structure traced by the {\chn \N2Dp}, as well as four young stellar objects. The \N2Dp filament is wider in the north (where it coincides with continuum dust emission detected in our ALMA observations), and in the southern part, the structure narrows and bends towards the east. 
The four YSOs in this region have been identified as Class I and Class II sources (which are more evolved than the sources in the other two regions to the south). Three of the sources coincide with our detected continuum emission. Yet, unlike the other two regions to the south, none of the YSOs overlap with the 
\N2Dp emission (as expected for more evolved YSOs). The CO(2-1) data (not shown here) reveal several outflows in this region, extending at least 0.1 pc which are very likely  interacting with the dense gas traced by the \N2Dp. The evolutionary stage of the sources, as well as the widespread outflow activity in {\chel the north} region suggests that this is the most evolved region of the three.

The main structure of the central region is a {\chn 85\arcsec (0.16\,pc)}-long \N2Dp filament. In the northern end of the filament we find three very young protostars: the known Class 0 protostars HOPS\,317 
\citep{2016ApJS..224....5F} 
and HH24mms \citep{1993A&A...272L...5C,1995MNRAS.274.1219W}, {\chtw and 
a new source about 10\arcsec\/ (4000 au) northwest of the other two sources which we name HH24mms-NW (see \autoref{fig3}b).}
Each of these protostars power their own compact outflow (extending only up to $\sim 0.05$ pc from its source) traced by high-velocity CO and H$_2$CO emission, which may be impacting the immediate surroundings of the protostar, but certainly not the filament as a whole.
The southern part of the central  \N2Dp filament (between declination of about -0:11:40 and -0:12:00) is coincident with two continuum emission peaks {\chtw (see \autoref{fig3}b)}. Compact high-velocity CO and H$_2$CO outflow emission within 0.02 pc of these continuum sources suggest that at least one of these harbors a very young protostar. 

In the southern region, the \N2Dp emission is composed of three condensations at around {\chel dec\,=\,-0:12:15},   a {\chn 105\arcsec}-long filament that extends to the south, and an isolated condensation in the southwestern edge of our map.  Both the eastern and central \N2Dp condensations at a declination of about -0:12:15 harbor a Class 0 protostar (HOPS 402  and HOPS 401, respectively, see  \autoref{fig1}b). The southern filament is home to two known Class 0 protostars, HOPS 316 and HOPS 358 (a.k.a., HH25mms), and to a number of previously undetected sources
which we will present in a forthcoming paper. South of the filament lies the flat-spectrum source HOPS 385 and the Class I source HOPS 315, which is coincident with the southwestern condensation. Several of the protostellar sources in the southern region power high-velocity outflow lobes seen in {\cht our (yet to be published)} CO ALMA data that extend about 0.1 pc from their driving source and are {\cht clearly} interacting with the filament. 


The elongated structures in these three regions have lengths ranging from 0.13\,pc to 0.20\,pc, and widths from about 0.01 to 0.04\,pc. Therefore, these structures are significantly smaller than the typical star-forming filaments observed in dust continuum with the Herschel Space Observatory, which have lengths of about one to several tens of pc  and widths of about $0.1$\,pc \citep{2014prpl.conf...27A, 2019A&A...621A..42A}.
The filamentary structures we detect are more similar in scale to the significantly narrower filaments  mapped in the Orion A molecular cloud using high resolution molecular line observations, like the narrowest  C$^{18}$O filaments reported by \citet{2019A&A...623A.142S}, and the so-called fibers observed by \citet{2018A&A...610A..77H} using N$_2$H$^+$.

In this paper, we will focus on the central region because of its {\chel comparative} simplicity, significantly lower  stellar feedback activity, and the relative young age of the filament and the sources in it. The quiescent central region is a good laboratory for studying the initial conditions for filament and core formation, while the south and north regions, which are more evolved and active, are more appropriate for studying how stellar feedback impacts cluster star formation in a filament (which  will be discussed in detail in a future paper).

To highlight the structure of the filaments, we plot the peak intensity of \N2Dp emission (Moment 8) in \autoref{fig1}b.  The central filament shows its brightest peak intensity in the northern half. However, close to the northern end of the central filament we can see a clear attenuation of \N2Dp emission {\chni at} the position of the highest intensity continuum peak {\chtw (HH24mms)}.   This is possibly due to the increase in temperature around this protostellar source as \N2Dp mainly traces low temperature regions \citep{2007A&A...467..179P,2015ApJ...804...98K,2018ApJ...867...94K}. 
 {\chtw South} of HH24mms, the filament shows local maxima in the \N2Dp moment 8 maps at the edges of the filament, around declination -0:11:00. 
Further south, at around declination -0:11:20,
the central filament is significantly narrower and the intensity profile is centrally peaked. 

{\chtw In \autoref{fig2} we present a velocity map derived from the 
 \N2Dp emission 
 of the entire region, and in \autoref{fig3} we zoom in on the central filament. 
 To produce these maps and those of  velocity dispersion (also shown in \autoref{fig2} and \ref{fig3}), we 
  modeled the brightest 25 hyperfine lines of  the N$_2$D$^+$(3-2) transition listed in Table 2 of \citet{2001ApJ...551L.193G} with 25 Gaussians, using the known relative frequencies and intensities of the hyperfine lines, and assuming the emission is optically thin.} 

{\chni In Figures 2a and 3a,   it can be seen that across approximately the entire length of the central filament there is an east-west velocity gradient. 
In addition, the  north  part  of  the  southern  filament  appears  to show a velocity gradient similar to that of the central filament (see Figure 2a).  It is possible that these two filaments were once connected and their kinematics had the same origin. The protostar HOPS 401 lies between these two filaments and its formation could have resulted in the current disconnect between the central and southern structures.  Although our observations cannot confirm this scenario, what is  apparent from our data is that these two filaments are now disconnected, and as such we will assume they are independent structures.

The velocity in the central filament gradient is purely along its minor axis  with no detectable gradient along the length (major axis) of the filament. 
This implies that either there is little or almost no gas following along the filament or the filament lies mostly on the plane of sky. From the channel maps shown in \autoref{fig4}, we can see the position of the filament changes from east to west as a whole as the $V_{lsr}$ increases, with no noticeable velocity structure alone the length of the filament.  As most filaments show some velocity structure along their major axis \citep[e.g.,][]{2014prpl.conf...27A}, this suggests that the central \N2Dp filament has little or almost no inclination with respect to the plane of the sky.  






\begin{figure*}[tbh]
\centering
\makebox[\textwidth]{\includegraphics[width=\textwidth]{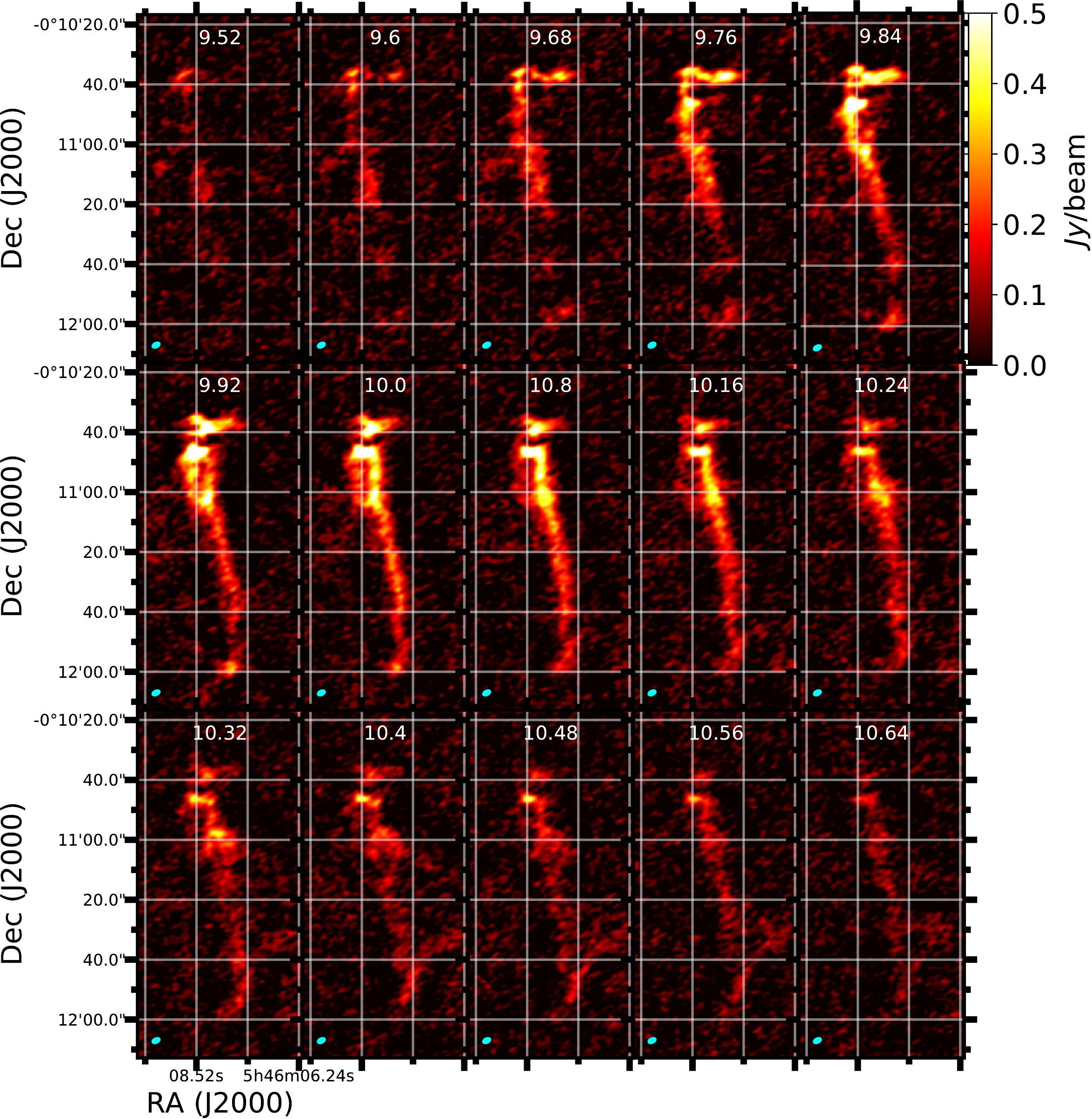}}
\caption{Channel maps of the \N2Dp ($J = 3-2$)
emission of the central filament. The  $V_{lsr}$ of each channel is shown at the top of each panel. {\chtw The cyan ellipse in the lower left corner of each panel shows the beam size of the \N2Dp map.}
}
\label{fig4}
\end{figure*}





{\chth In \autoref{fig3}b, we show the \N2Dp linewidth map. It can be seen that the linewidth mostly ranges between 0.15 and 0.35 km s$^{-1}$. Local maxima coincide, for the most part, with continuum peaks which trace pre-stellar and protostellar envelopes.
In these regions the increase in linewidth is likely due to unresolved motions surrounding these sources, such as outflows and infall. 
 Between -0:11:00 and -0:11:20 
 we also see a local maximum of linewidth along the spine  and toward the eastern edge of the filament, and south of -0:11:20 the linewidth is higher toward the edges. In summary, there is no ordered  structure in the linewidth map across the entire filament as there is in the velocity map.
}

In principle we could use the \N2Dp emission to derive the \N2Dp column density and obtain an estimate of the total filament mass. However, 
vast variations in the deuterium to hydrogen abundance ratio in clouds and cores  can yield  \N2Dp abundance ratios that range a few orders of magnitude \citep{2002ApJ...565..344C,2006ApJ...647.1106L,2015MNRAS.446.1245L}. 
Therefore, mass estimates from measurements of the \N2Dp column density, assuming a constant \N2Dp to H$_2$ abundance ratio are  highly uncertain. Here, instead, {\chel we} use the 850\,$\micron$ Hershal-Plank dust emission map \citep[from][]{2014A&A...566A..45L} in concert with our ALMA data to estimate the  H$_2$ column density and \N2Dp abundance  of the central filament (see \autoref{sec:appendix}).


To convert H$_2$ column density to number density, we assumed {\chtw a cylindrical filament and} that the {\chtw average} depth of the filament (the dimension along the line of sight) is equal to the {\chtw radius multiplied by $\frac{\pi}{2}$ (i.e., the area of a circle divided by its diameter)}. {\chtw The central filament has dimensions (as seen on the plane of the sky) of 13\as8 $\times$ 87\as6  (5540  $\times$ 16640\,AU),\footnote{{\chtw Note that the filament width varies (it becomes narrower towards the south, see \autoref{fig3}). For all the number density and mass calculations, we use the average width of 13\as8 (5540\,AU), estimated from the Moment 0 map where only emission above 4\,$\sigma$ is included.}}} we thus estimate a {\chtw peak and average} number density for {\chtw the central filament to be  approximately $6.6 \times10^{6}$\,cm$^{-3}$ and $1.6 \times10^{6}$\,cm$^{-3}$}, respectively. 
We sum the column density within the area of the filament to obtain the total filament mass of {\chtw 11.7} M$_\odot$, which corresponds to a linear mass of {\chtw 64.6} M$_\odot$ pc$^{-1}$. 
We estimate an approximate 50\% uncertainty in our calculation of the filament mass, which results from the error in the fit to the empirical relation between the N$_2$D$^+$ integrated intensity and the
H$_2$ column density (see Appendix A) 
and the range of mass estimates obtained when choosing a range of thresholds for the minimum signal-to-noise emission that is used for determining the mass.

{\chtw Assuming this is an isothermal filament with a temperature of 10\,K, we use Equation 59 in  \citet{1964ApJ...140.1056O}  to estimate the critical mass per unit length\footnote{The equation for  the critical mass per unit length is: $M(x)=\frac{2kT}{\mu m_p G}(1+\frac{1}{x^2})^{-1}$,
where $x$ is the dimensionless width shown in \citet{2014MNRAS.444.1775R}. In our case $x=3$ (See Sec.~4.1). The commonly used value in the literature of $M_{line,crit}\sim 16$ M$_\odot$ pc$^{-1}$ results from the assumption of much larger filament widths, with  
$x \rightarrow \infty$ \citep[e.g.][]{2014prpl.conf...27A}. 
}
(for a non-rotating filament) to be {\chtw $M_{line,crit}\sim 25.2$ M$_\odot$pc$^{-1}$}. This is lower than the estimated linear mass for the central filament, 
which indicates the filament is gravitational unstable. As we discuss 
below, the central filament in LBS23 is rotating and this 
should be considered in the stability analysis of this filament. That is what we do in the following section.


\section{Discussion}
\label{sec:discussion}

\begin{figure*}[!htp]
\centering
\subfloat{%
  \includegraphics[width=0.3\textwidth]{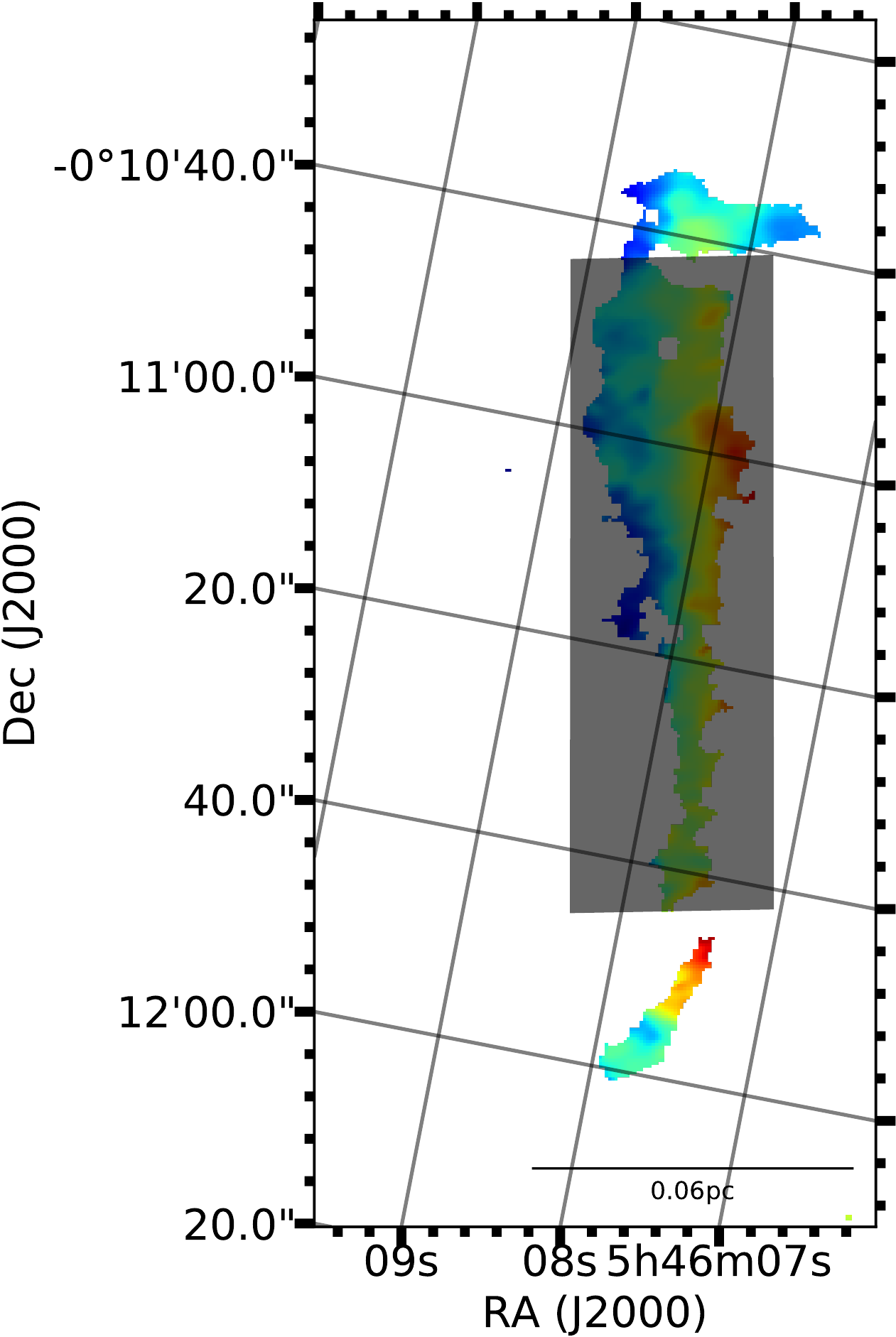}%
}\quad
\subfloat{%
  \includegraphics[width=0.65\textwidth]{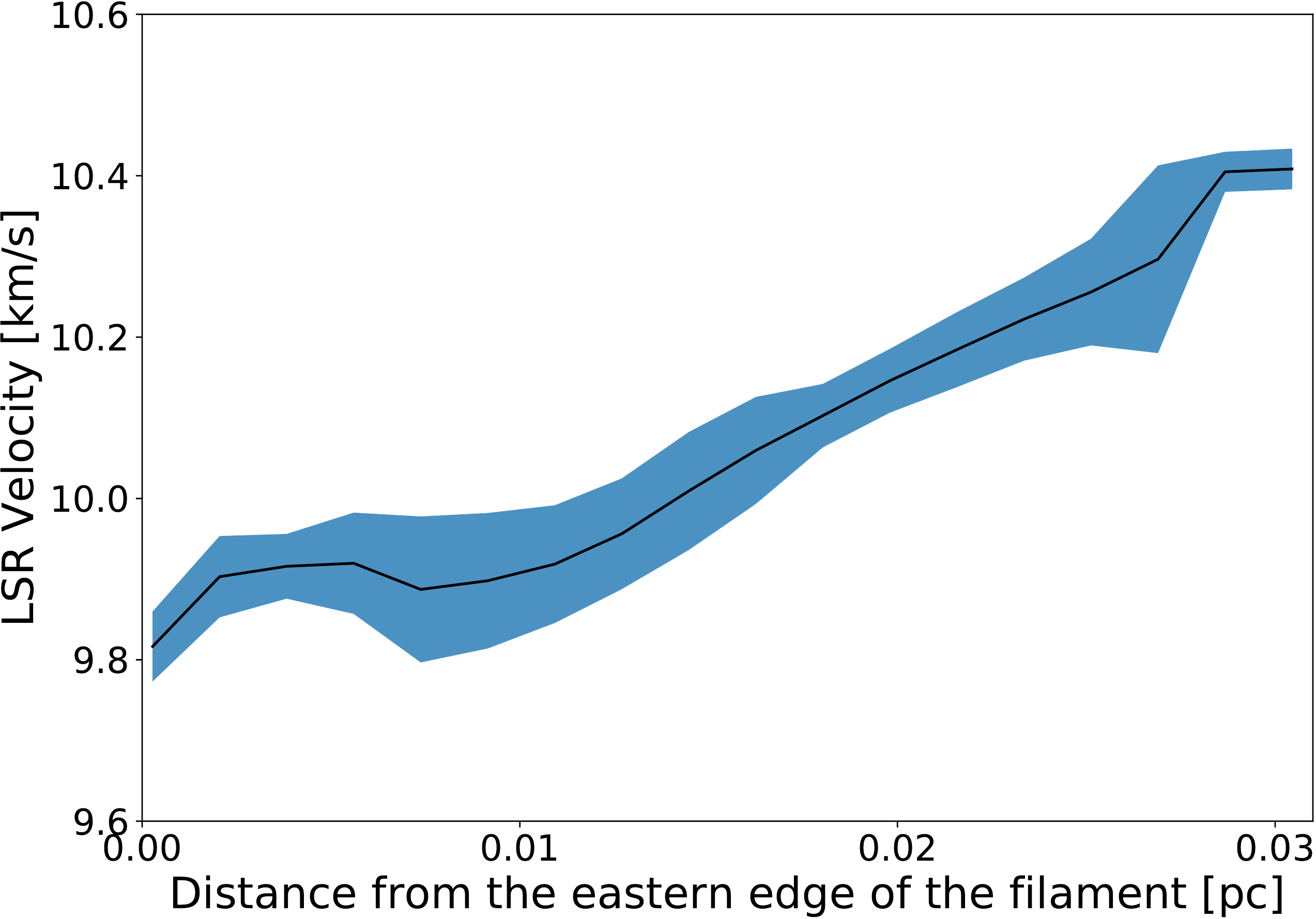}
}
\caption{{\chni \N2Dp  system velocity profile across the central filament. The left panel shows the velocity map (from Figure 2a)  and shows the area from where the average velocity profile along the filament minor axis, shown in the right panel, was obtained.
The figure has been rotated by 11\arcdeg \/ (north of east) in order to have the filament's long axis parallel to the y-axis of the plot.
{\chth The black line in the right panel represent the average velocity profile.}
  The blue region surrounding the black line in the right panel represents the standard deviation around the average. 
}}
\label{fig5}
\end{figure*}

\subsection{Rotation and its effect on filament dynamics}
\label{sec:rotation}
In \autoref{fig2}a, we observed a clear velocity gradient along the minor axis of the central filament. 
To further study this kinematic structure we plot the average velocity profile across the filament in {\chtw \autoref{fig5}.} {\chtw Here we see a clear increase in radial velocity from east to west across the central filament.} We interpret this gradient to be caused by rotation in the filament {\chtw (see \S.\ref{sec:why_rotation} for a discussion on this)},
{\chtw and use the  velocity profile to estimate the rotation velocity. In \autoref{fig5} we see that the velocity increases from about 9.8\,km\,s$^{-1}$ in the eastern edge of the filament to 10.4\,km\,s$^{-1}$ in the west.}
The rotation velocity is determined to be half of the total velocity range of the filament, or about  $0.3$\,km\,s$^{-1}$.
Dividing the presumed circumference of the cylinder by the rotation velocity, we estimate the rotation period {\chtw ($T=\frac{2\pi r}{v}$, where $\frac{v}{r} = 20 \pm 1 $ km\,s\,pc$^{-1}$)} {\chtw to be $3.1\times10^5$\,years, which corresponds to an angular frequency ($\omega$) of $6.5\pm 0.3 \times 10^{-13}$\,rad\,s$^{-1}$}.

To further analyze the dynamics of the filament, we compare our results with the theoretical study of \citet{2014MNRAS.444.1775R}, which considers a rotating filament model in hydro-static equilibrium.
Following Recchi et al., we calculate the normalized angular frequency\footnote{
$\Omega= \omega \sqrt{2/\pi G \rho_0}$, 
where $\rho_0$ is the central density.} ($\Omega$)
for the {\chtw central filament to be 0.36.} To obtain this value 
we used a central density ($\rho_0$) of about  {\chtw $3.0\times 10^{-17}$\,g\,cm$^{-3}$} (obtained from the estimate of the number density  given above).  
If the centrifugal force balances  the gravitational force (i.e., $\Omega=2.0$), then the filament would be in Keplerian rotation and would have a constant density profile \citep{1978PASJ...30...39I,2014MNRAS.444.1775R}. {\chni In our case, $\Omega<2.0$ and thus the centrifugal force is less than the gravitational force. 
Using our estimate of the central density and assuming a temperature of 10\,K, we estimate the dimensionless truncation radius ($x$)\footnote{$x =r/\sqrt{\frac{2kT_0}{\pi G \rho_0 \mu m_H}}$, where $T_0$ is the central temperature  \citep[see][]{2014MNRAS.444.1775R}.}.
Assuming the filament is isothermal, {\chn and using the same theoretical calculation as that used to obtain the values of Table 1 in \citet{2014MNRAS.444.1775R}}, we find the critical linear mass
($M_{line,crit}$), for an isothermal rotating filament with truncation radius
of {\chtw $x=3.0$ 
and normalized angular frequency of  $\sim0.36$
is 29.0}} M$_\odot$\,pc$^{-1}$.
This value is slightly larger than 
the critical linear mass  for an isothermal 10\,K non-rotating cylinder \citep{1964ApJ...140.1056O,1997ApJ...480..681I}. 
 Our results show that the central filament 
 has a  linear mass that is significantly higher than the critical linear mass (even after taking rotation into account), and thus the filament is not stable against collapse.
 This is consistent with the evidence of fragmentation observed in the continuum emission and protostellar activity in both the northern {\chtw and southern edge of the central filament} (see Sec. \ref{sec:results}).

 To further understand the importance of rotation in the filament's dynamics, we calculate $\beta_{rot}$, the ratio between rotational kinetic energy to gravitational energy (see \autoref{sec:appendixB} for a detail description of how we calculated these quantities). We obtain a value of $\beta_{rot} \sim 0.04$ for the central filament in LBS\,23. This value is consistent with our result that the central filament is not stable against
 collapse, even when considering rotation. Our estimate of $\beta_{rot}$ is similar to the average value obtained for cores by \citet{1993ApJ...406..528G} and in agreement to the estimate of $\beta_{rot}$ for the cores and envelopes of similar size as our filament listed by \citet{2013ApJ...768..110C}.
 Even though the rotation in the filament is not enough to prevent collapse, it may still trigger the formation of instabilities (e.g., fragmentation, bars, rings). For example, the numerical simulations by \citet{1999ApJ...520..744B} and \citet{2011MNRAS.417.2036B} show that cores with $\beta_{rot} \geq 0.01$ can form  a dense flattened structures that may then fragment. 
 These simulations modeled rotating molecular cloud cores, thus their results may not be entirely applicable to filaments. Simulations of collapsing rotating filaments should be conducted 
 to determine the importance of rotation in triggering fragmentation and the formation of other instabilities in filaments.

\begin{figure*}[tbh]
\centering
\makebox[\textwidth]{\includegraphics[width=\textwidth]{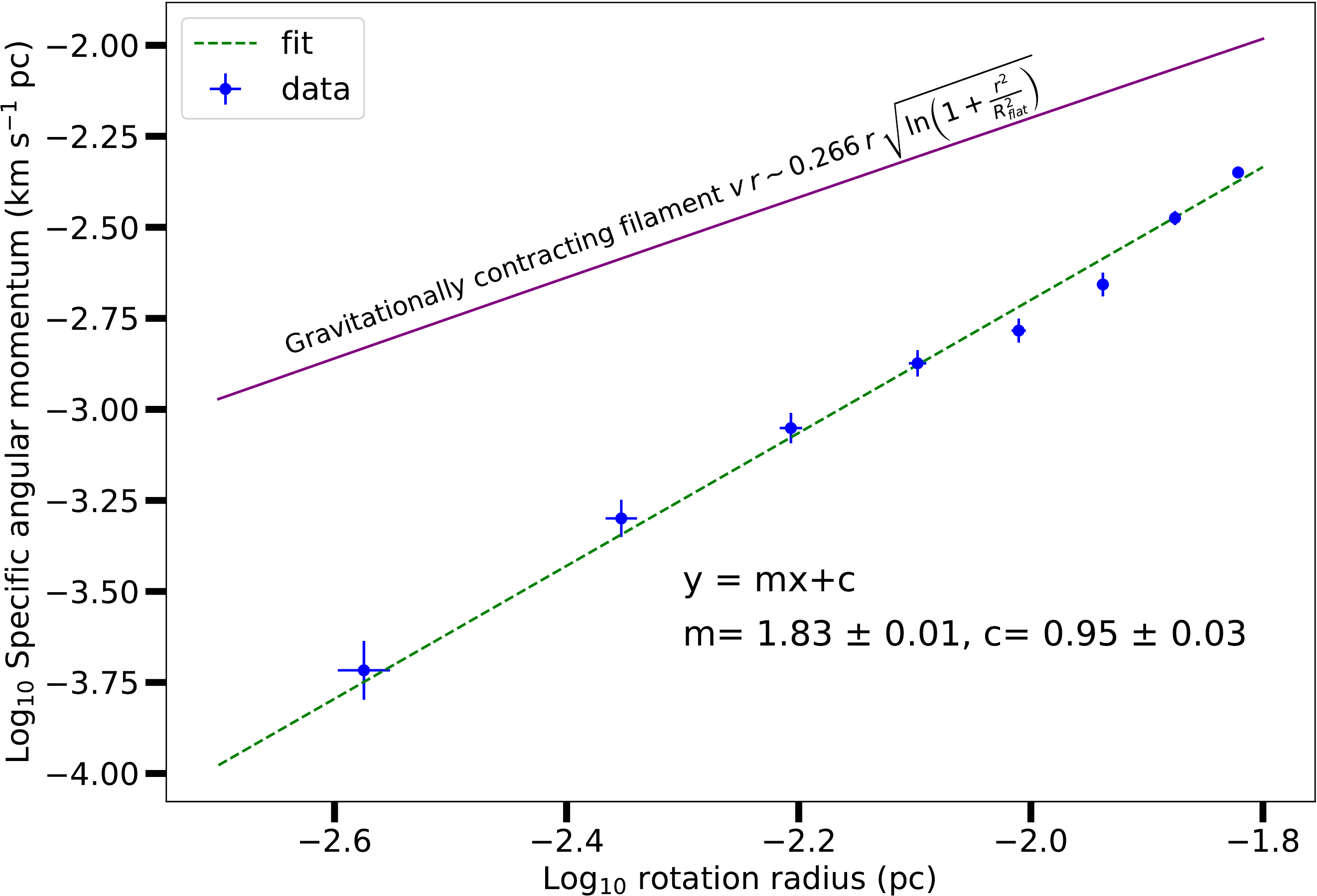}}
\caption{ Radial profile of the specific angular momentum ( $j = r \times V_{rot}$) for the N$_2$D$^+$ filament,  derived from the velocity profile shown in \autoref{fig5}. The blue points represent the average rotational velocity obtained by averaging the velocities at similar distances from the filament center on the eastern (blue-shifted) side and 
the western (red-shifted) side of the filament. 
The dotted green line shows the best-fit  line to the data in log-log space. The purple line represents the expected $r \times v$ profile for a gravitationally contracting filament (see Sec.~4.6)}.
\label{fig7}
\end{figure*}

\subsection{{\chtw Specific angular momentum profile of filament}}

{\chel Various observational studies have measured the specific angular momentum in cores and envelopes, with scales from  $\sim 0.01$pc to about $0.1$\,pc \citep{1998ApJ...504..223G,2002ApJ...572..238C,2016PASJ...68...24T} and in early Class 0/I disks (with scales $\leq$100\,AU) \citep{1997ApJ...488..317O,2007ApJ...669.1058C,2012ApJ...748...16T,2013ApJ...765...85K,2015ApJ...812..129Y}. Based on these measurements, it has been proposed that specific angular momentum is conserved from scales of the inner envelope (a few $10^3$ au) to disk scales ($\sim 10-100$au)
 \citep{1997ApJ...488..317O,2013EAS....62...25B,2020A&A...637A..92G}. In order to understand how angular momentum varies (or is conversed) at various scales, it is essential to obtain specific angular momentum measurements of different structures of different sizes (i.e., filaments, cores, envelopes and disks). In this work we concentrate on the filament scales.

{\cht We} measure the specific angular momentum of the filament {\cht by} assuming a rotating cylinder model. The specific angular momentum {\chtw $J=L/M$} can be expressed as:
\begin{gather}
    {\chtw J=\frac{L}{M}=\frac{I\omega}{M}}.
\end{gather}

Using the angular frequency $\omega$ and radius $r$ measured in \S.\ref{sec:rotation} {\chtw and the moment of inertia ($I$) calculated in \autoref{sec:appendixB}}, we find the total specific angular momentum for the {\chtw central filament is $\sim 4 \pm 2 \times 10^{20}$\,cm$^{2}$\,s$^{-1}$.}

\citet{2016PASJ...68...24T} measured the total specific angular momentum of 27 N$_2$H$^{+}$ cores in Orion A using the Nobeyama 45\,m radio telescope, and found it ranges between about $ 10^{20}$ to  $10^{21}$\,cm$^{2}$\,s$^{-1}$ (see their Table 2).
These cores have a typical size of approximately 0.04\,pc to about 0.1pc which is comparable to our \N2Dp filament width of about 0.04\,pc.
N$_2$H$^{+}$ and N$_2$D$^{+}$ are both high-density tracers, and in the cold pre or proto-stellar cores where CO freeze onto dust grains, the ratio N$_2$D$^{+}$/N$_2$H$^{+} $ is about  0.24 \citep{2002ApJ...565..344C}. The near-unity abundance ratio, as well as the similarity of structures in maps of young cores using these species \citep[e.g.][]{2013ApJ...765...18T} suggest that both species trace somewhat similar density regimes. The similarity in the specific angular momentum between cores and our filament may suggest that angular momentum of cores is linked to the rotation of small filaments.


{\chtw Our observations have enough resolution to be able to determine how the specific angular momentum varies with distance from the center of the filament.
 In \autoref{fig7} we plot the derived specific angular momentum profile ($j(r)$) for the central N$_2$D$^+$ filament. 
 The measured specific angular momentum as a function of radius is given by $j(r) = r \times V_{rot}$, where $r$ is the radius from the filament center (which in \autoref{fig5} corresponds to 0.15 pc from the eastern edge of the filament), 
 and $V_{rot}$ is the rotational velocity around the filament's rotational axis. In our case $V_{rot} = |V(r) - V_c|$, where $V(r)$ is the $V_{lsr}$ at radius $r$ from the center of the filament, given by the plot in \autoref{fig5}, and $V_c$ is the $V_{lsr}$ at $r=0$, which is  10.05\,km\,s$^{-1}$ . 

A fit to the data (using $j(r) \propto r^{\alpha}$) gives an exponent of  
 $1.83\, \pm \,0.01$ (see \autoref{fig7}). This is consistent with the relation of total specific angular momentum as a function of core radius ($J \propto R^{1.6\pm 0.2}$) for a sample of cores derived by \citet{1993ApJ...406..528G}, and the results of more recent work 
 which obtain the average specific angular momentum profile of a sample of young protostars \citep{2019ApJ...882..103P,2020A&A...637A..92G}.

{\chtw 
The similarity in 
the angular momentum profile of our filament and that of the dense cores and envelopes 
 suggests that the  angular momentum of the dense circumstellar environments  may be linked to (and even inherited from) the rotation of small 
 filaments. 
One (naive) first step to determine if such a link exist in  the central filament in LBS23 is to compare the rotational axis of the envelopes surrounding the protostars in the filament with the rotational axis of the filament. The resolution of our observations is not enough to obtain a reliable rotational axis from the envelope emission. Instead we use the outflow axis as a proxy for the envelope/disk rotation axis.  

In \autoref{fig3} we plot the outflow axes of the three protostars in the central filament of LBS23 with reliable outflow detection.
Even though the outflow axes are clearly not aligned with the filament rotation axis, it would be too premature to conclude that there is no link between the filament and envelopes from this simple comparison. It could be that although the outflows are tracing the spin axis of the circumstellar disk, it is not the same as that of the envelope \citep[e.g.,][]{2018MNRAS.475.5618B}. 
In addition, given the small projected separation between the three protostars in the northern end of the filament 
(less than 4000 au)
it is very likely that 
they are part of a (hierarchical) triple system \citep[e.g.,][]{2013ApJ...768..110C,2016ApJ...818...73T}. If that is the case, then the  total angular momentum of the system, which consists of the envelope spin and the orbital angular momentum of all members,
is the quantity that should be associated to the specific angular momentum of the filament and not  the spin of the individual members.

Studies suggest that the observed rotation in cores and envelopes, with a power-law dependence with an index of about 1.6, is produced by the turbulence cascade of their parent molecular cloud \citep{2018ApJ...865...34C,2020A&A...637A..92G}. Our filament shows a similar dependence and thus it is tempting to suggest that the velocity gradient seen along the minor axis of  our filament is acquired from turbulence as well. In our filament, this power-law dependence is seen to continue down to our resolution level of about several $10^2$ au. This is different from the observed flattening of the 
specific angular momentum as a function of radius (or size) at scales of a few $10^3$
au in a sample of various young stellar systems \citep{1997ApJ...488..317O,2013EAS....62...25B,2014prpl.conf..173L,2020A&A...637A..92G}. 
This flattening is thought to indicate the scale for dynamical collapse, where  angular momentum is conserved.

In the traditional two-step scenario where thermally supercritical filaments form first and cores then form by gravitational fragmentation  \citep{1964ApJ...140.1056O,1992ApJ...388..392I,1997ApJ...480..681I}, we would expect a clear transition between the  angular momentum profile of the filament (inherited from the cloud turbulence) with a power-law-dependence and a flattening of the angular momentum profile at the core scales, where   gravity dominates and angular momentum is conserved. In contrast to this scenario, our observed filament shows a profile that is consistent with specific angular momentum profile set by turbulence all the way down to a few hundred au (scales that are even smaller than the size of the triple system at the northern end of the filament). Our observations are thus more consistent with the scenario in which  filament and cores develop simultaneously due to the multi-scale growth by nonlinear perturbation generated by turbulence
\citep[see numerical simulation studies by][]{2011ApJ...729..120G,2015ApJ...806...31G,2014ApJ...785...69C,2015ApJ...810..126C,2014ApJ...791..124G,2014ApJ...789...37V}.} 
In this picture the  initial angular momentum of a filament and the cores inside it are first acquired from the ambient turbulence.  Subsequent gravitational interactions among dense condensations may redistribute the angular momentum of individual cores and envelopes \citep[as suggested by][]{2019ApJ...876...33K}. This could also explain the difference in the outflow axes and the filament rotation axis.}

\subsection{{\chtw  Transonic turbulence in the filament}}

\begin{table}
\setlength{\tabcolsep}{4.6pt} 
\caption{{\chtw  Estimated energies for the central filament} }            
\label{table:1}      
\centering                          
\begin{tabular}{c  c c}        
\hline\hline                 
{\chtw Energy Type} & {\chtw Value (erg)}& {\chtw Ratio ($1/E_G$)} \\
\hline                        
{\chtw Gravitational Energy ($E_G$)} & {\chtw $7.1\times10^{43}$}   & {\chtw 1.0} \\ 
{\chtw   Turbulence Energy ($E_{\sigma}$)} & {\chtw $9.7\times10^{42}$}   & {\chtw 0.14} \\      
{\chtw   Rotational Energy ($E_{R}$)}  & {\chtw $3.0\times10^{42}$}   & {\chtw 0.05} \\
{\chth   Magnetic Field Energy ($E_{B}$)}  & {\chth              {\chft $3.4\times10^{43}$} }  &  {\chft  0.48 } \\
\hline                                   
\end{tabular}
\end{table}

{\chtw One general way to understand the properties of a filament is to characterize its turbulence. Previous observations of cores by \citet{1998ApJ...504..223G} have shown that  medium-density tracers such as C$^{18}$O  show supersonic velocity dispersion while denser tracers such as NH$_3$ shows velocity width comparable to the thermal line width. \citet{2010ApJ...712L.116P} used NH$_3$ as a high-density tracer to study the B5 region in Perseus. They found a $\sim 0.1$\,pc-wide filamentary region with sub-sonic turbulence  surrounded by supersonic turbulence. The subsonic coherent cores  mark the point where most of the turbulence decays and the gas is ready to form protostars \citep{1998ApJ...504..223G,2002ApJ...572..238C,2013ASPC..476...95A}. 

We measured the non-thermal motion in our filament using the following equation:
   \begin{gather}
      \sigma_{NT}  = \sqrt[]{{\sigma_v}^2-\frac{k_B\,T_g}{m_{N_2D^{+}}}}
          \label{eq:veldis}       
   \end{gather}
where $\sigma_v $ is the observed velocity dispersion shown in \autoref{fig3}b, $k_B$ is the Boltzmann constant, $T_g$ is the gas temperature, and $m_{N_2D^+}$ is the mass of the \N2Dp molecule. From \autoref{fig3}b, we can see the typical velocity dispersion along the line of sight ranges from about 0.15 to 0.35 km\,s$^{-1}$. The velocity dispersion is greater in the regions with evidence of protostellar activity, close to the positions of the continuum sources in the northern and southern edges of the filament.
We find the average velocity dispersion
within 
the area (chosen to avoid outflow ``contamination'') south of HH24mms and north of the continuum peaks in the southern end  of the central filament 
is 0.20 km\,s$^{-1}$.
Assuming a temperature of 10\,K, the sound speed ($c_s = \sqrt{k_B T_g/\mu m_H}$, where $\mu=2.33$ is the mean molecular weight, and $m_H$ is the mass of the hydrogen atom) is around 0.19\,km\,s$^{-1}$. Thus, the average non-thermal velocity component is estimated to be $\sim 0.19$\,km\,s$^{-1}$, resulting in an average Mach number ($M_s$) of  1.0 for the central filament. This is in contrast with the sub-sonic turbulence ($M_s < 1$) that is generally expected for dense structures with scales less than 0.1 pc  and the picture of coherent cores \citep{1998ApJ...504..223G,2002ApJ...572..238C}. However, our finding is consistent with recent observations of a few star-forming filaments/fibers with  transonic  turbulence \citep{2016ApJ...833..204F,2017A&A...606A.123H}. 
The existence of young protostars in a transonic filament, as it is the case in the central filament of LBS23, suggests that star formation can occur  before the turbulence fully decay from supersonic to subsonic.

In order to assess the relative importance of turbulence, we estimated the turbulence energy of the central filament using: \begin{gather}
E_{\sigma} = M\sigma_{NT}^2 
\end{gather}
where $\sigma_{NT}$
is the non-thermal velocity dispersion, and $M$ 
is the mass of the central filament \citep{2000MNRAS.311...85F}. Using our estimates for these two quantities we obtain a turbulence energy of $9.7 \times 10^{42}$ erg, with a ratio of turbulence\textbf{} to gravitational energy of 0.14 (see \autoref{table:1}). 
 A significantly lower value of the turbulence energy compared to the gravitational energy should be expected in a dense region where protostars are forming. 
 
{\chth The turbulence energy  in the central filament is about a factor of three larger than the rotational energy ($E_{rot}$). Protostellar cores and envelopes with velocity gradients indicative of rotation generally have rotational energy that are significantly smaller than the turbulence energy. 
This can be clearly seen in the results of  \citet{2007ApJ...669.1058C} who studied a sample of protostellar envelopes, traced by the N$_2$H$^+$ emission.  
In addition, the study of 
\citet{2016PASJ...68...24T} found  that for a typical core in Orion the average velocity gradient, presumably due to rotation, is $2.4 \pm 1.6$\,km\,s$^{-1}$\,pc$^{-1}$ and the average core diameter is $0.08 \pm 0.03$\,pc. Therefore, the average rotation velocity is estimated to be $\sim 0.19$\,km\,s$^{-1}$. A large survey of 71 cold cores in Taurus, California and Perseus  found an average non-thermal velocity width of $\sim 0.3$\,km\,s$^{-1}$ \citep{2013ApJS..209...37M}. Thus, we expect that cores will have   $E_{\sigma}/E_{rot} \approx (\frac{0.3}{0.19})^2\approx2.5$; consistent with our results for the central filament in LBS 23. 
Even thought the rotational energy is relatively small, rotation can still have an impact on the kinematics and structure of the system (see Sec.~4.1).
}

\begin{figure*}[!htp]
\centering
\subfloat{%
  \includegraphics[width=0.30\textwidth]{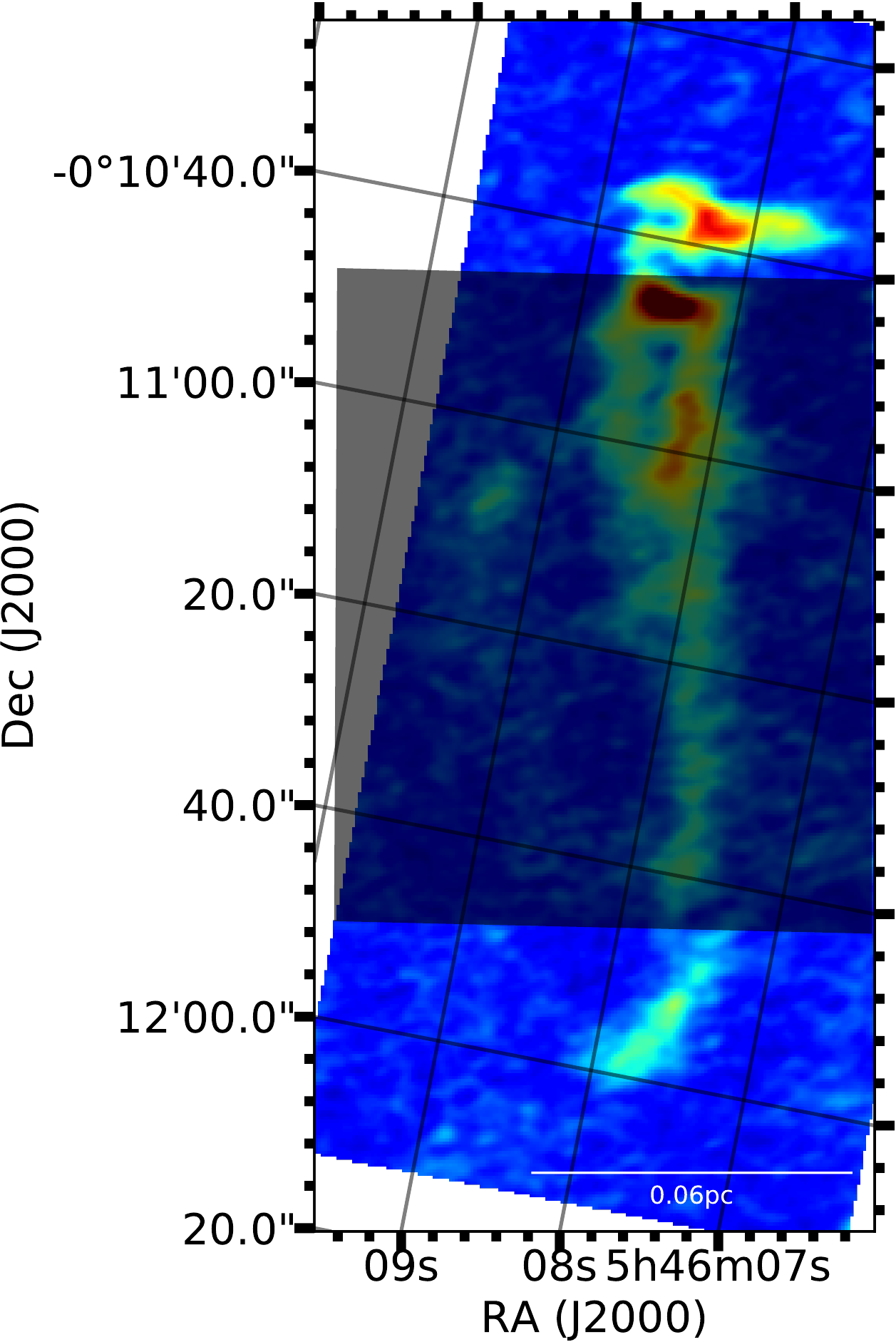}%
}\quad
\subfloat{%
  \includegraphics[width=0.65\textwidth]{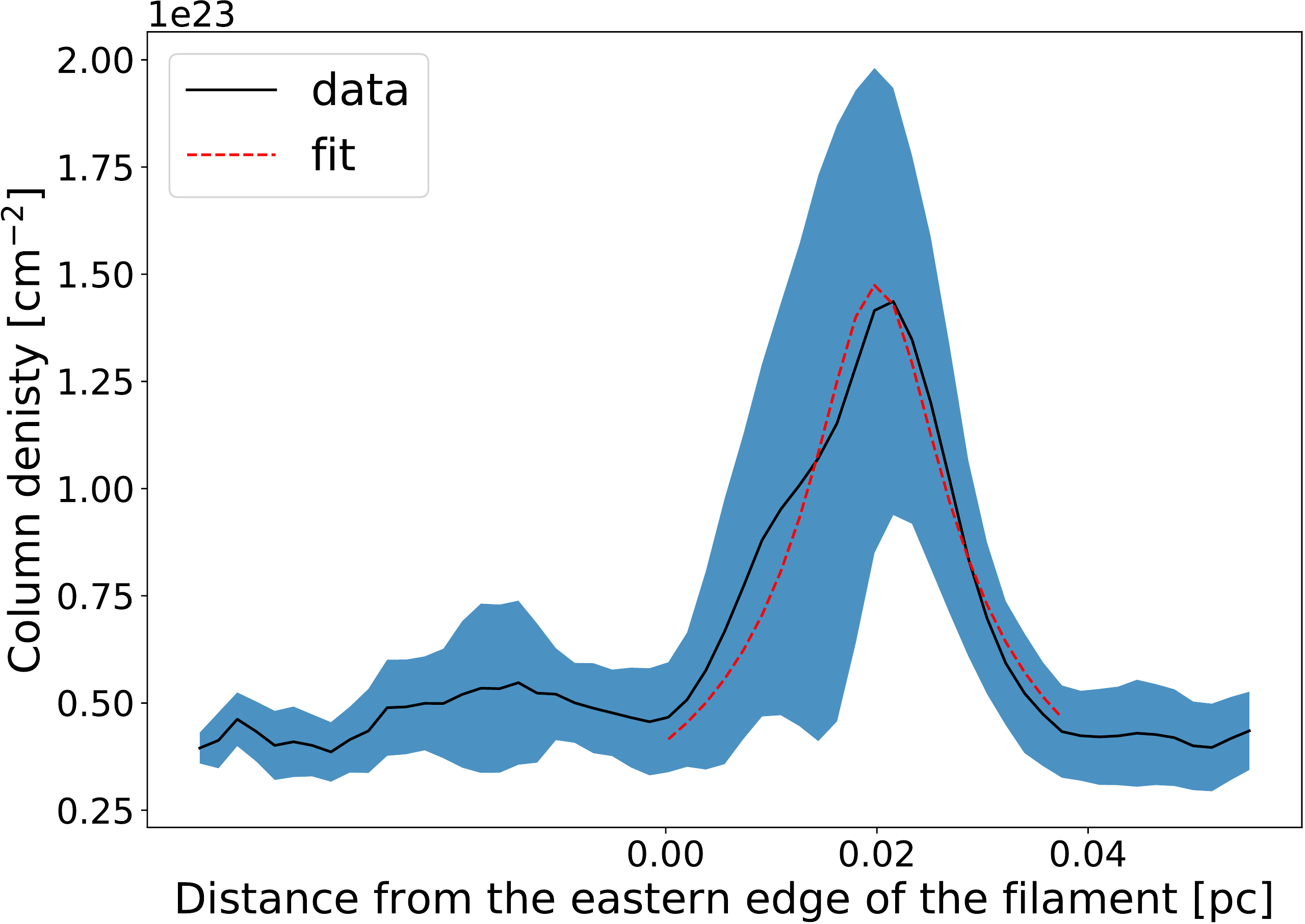}
}
\caption{{\chtw H$_2$} column density profile of the  central filament in LBS23. The  figure in the left panel shows the \N2Dp column density map (see Appendix B). The 0.18 pc long shaded area shows the region from where we obtain the average column density profile. The figure has been rotated by 11\arcdeg \/ (north of east) in order to have the filament's long axis parallel to the y-axis of the plot.
The black line in the right panel shows  the  average \N2Dp column density, and  the blue shade surrounding it represents the standard deviation around the average. 
The dotted red line represents the best-fit to the Plummer profile (see Eq.~4, and Sec.~4.4).}
\label{fig6}
\end{figure*}

We investigate whether this level of turbulence can be maintained in the filament and how it will evolve. To do this we first  estimate the turbulence dissipation rate, given by $L_{\sigma}=E_{\sigma}/t_{diss}$ \citep[e.g.,][]{2011ApJ...742..105A}. The turbulence dissipation timescale ($t_{diss}$) is given by $t_{diss}=\eta t_{ff}$, where  $t_{ff}$ is the free-fall time and $\eta$ ranges between $\sim1-10$ \citep{1989ApJ...345..782M,1999ApJ...524..169M}. \citet{2012ApJ...756..145P} estimated the uniform cylinder collapse time ($t_{ff,cylinder}$) to be $t_{ff,cylinder}=\sqrt{\frac{2}{3}}At_{ff,sphere}$. Here $A$ is the aspect ratio of the cylinder, which for our case is approximately 6, and $t_{ff,sphere}$ is the classical free-fall timescale of a uniform-density
sphere with the same volume density as the cylinder. Adopting $\eta=5$ and the average H$_2$ number density of the filament  $1.56\times10^6$\,cm$^{-3}$ (see Sec. \ref{sec:results}), we find the free-fall time for our filament to be  $1.9\times10^5$\,yr, which leads to  a turbulence dissipation rate of $3.3\times10^{29}$\,erg\,s$^{-1}$ for the central filament. \citet{2020ApJ...896...11F} observed 45 protostellar outflows in Orion A, and found the kinetic energy injection rates of
outflows are comparable to the turbulent dissipation rate. 
The energy ejection rate 
of outflows in Orion A ranges from about $10^{30}$  to a few  $10^{32}$ erg\,s$^{-1}$ \citep{2020ApJ...896...11F}. If we assume that the four
outflows in the filament have similar  energy injection rates to those in Orion A, then these protostellar outflows have more than enough power to maintain the turbulence in the filament, even if we were to assume a low efficiency in the coupling between
outflow energy and filament turbulence.
The excess outflow energy ejection rate could eventually increase the turbulence energy in the filament and prevent it from further collapse.




\subsection{{\cht Filament density profile}}
Another property that is commonly determined from observations is the filament density profile, as it may provide information on the dynamical stability of the filament and its formation.
In \autoref{fig6} 
we show the column density profile perpendicular to the {\chtw central} filament
averaged over a {\chtw 0.18\,pc} long region along the filament length. 
We then fit the average column density profile with a Plummer-like profile:  
\begin{gather}
    N(r)=   \frac{N_0}{\Bigg[(1+(\frac{r}{R_{flat}})^{2})\Bigg]^{(p-1)/2}}.
\end{gather}

{\chft The fit gives values for the power law index ($p$) of  $2.1 \pm 0.2$, the central (peak) column density ($N_0$) of $1.48 \pm 0.09 \times 10^{23}$ cm$^{-2}$, and the radius of the inner flat region ($R_{flat}$) of 
$0.006 \pm 0.002$ pc.}
The value of $p$ we obtain for
the central filament is significantly lower than the steep power law index  of $p = 4$ expected for an isothermal non-rotating cylinder in hydrostatic equilibrium  \citep{1964ApJ...140.1056O}. On the other hand, our result for the central filament in LBS23 is consistent with observations of filaments of various sizes \citep{2011A&A...529L...6A,2019A&A...621A..42A,2013A&A...550A..38P,2016A&A...586A..27K} as well as   hydrodynamic  and MHD simulations \citep{2014ApJ...791..124G,2014MNRAS.445.2900S,2016MNRAS.457..375F} of clouds which show that filament column density profiles can be well-fitted  with a Plummer-like profile with $p \sim 2$.
\citet{2016MNRAS.457..375F} argues that such density profile can be explained by filaments formed in the collision of two planar shocks in a turbulent medium, as the structure formed from this collision is expected to have a density profile that scales as $r^{-2}$, which corresponds to $p=2$.

\begin{figure}
\centering
\includegraphics[width=\hsize]{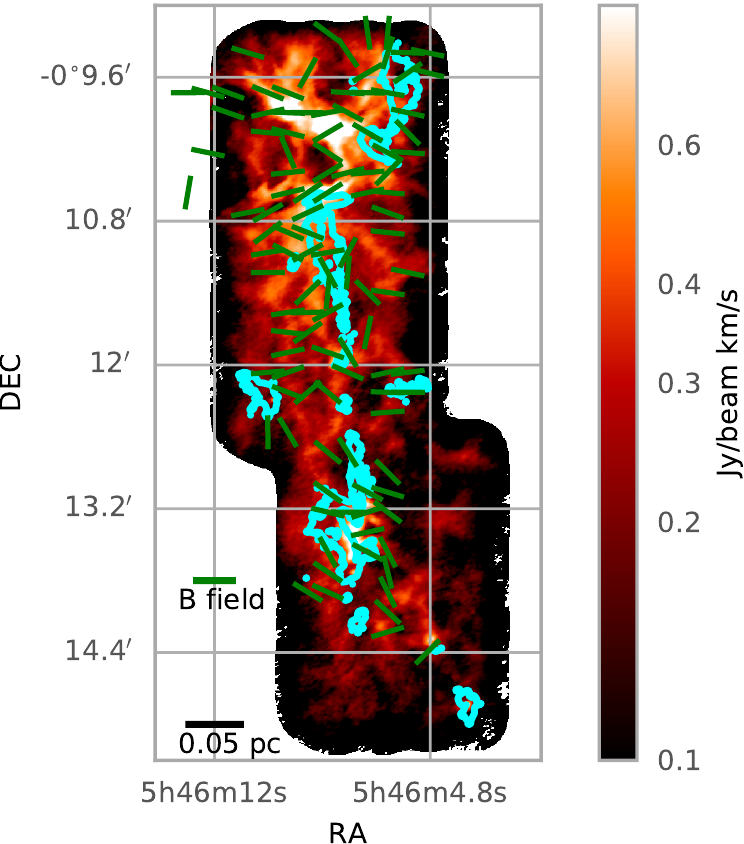}
\caption{ Plane of sky magnetic field direction inferred from archival SCUBA JCMT polarization data \citep{2009ApJS..182..143M}
superposed  {\chft on 
an integrated intensity map of the C$^{18}$O(2-1) emission of the region (color map) and 
the
N$_2$D$^+$ peak intensity  {\chth (contours)} from \autoref{fig1}b. The dark green lines represent  the direction of the B-field derived from the polarization  data. All lines are the same length regardless of the polarization percentage.} 
}
\label{fig8}
\end{figure}

\subsection{{\chtw Magnetic fields in the LBS\,23 filament}}

For a rotating system, magnetic braking effect can quickly slow down rotation and this has been found to be one of the challenges in disk formation theories \citep{2011ApJ...738..180L}. If the rotating filament is threaded by magnetic fields,  the magnetic tension could slow down the rotation. To study the magnetic fields in the LBS\,23 filament, we use the archival $850$\,$\mu$m SCUBA polarimeter data (with a beam of $\sim 20"$) from the James Clerk Maxwell Telescope (JCMT) presented by
\citet{2009ApJS..182..143M}. Dust grains are expected to have their long axes perpendicular to the magnetic field direction, resulting in polarized thermal emission from anisotropic aligned dust grains \citep[e.g.,][]{2007JQSRT.106..225L}.  In \autoref{fig8} we plot the plane of sky magnetic field direction 
in the LBS\,23 filament, 
obtained by rotating the polarization vectors by $90^{\circ}$
from the SCUBA JCMT polarization data. The figure shows that the plane of sky magnetic field is mostly perpendicular to the filament direction. 

We  use the dispersion in the distribution of polarization angles to derive the strength of the component of the magnetic field on the plane of the sky ($B_{pos}$) in the LBS\,23 region. For this, we use the 
Chandrasekhar-Fermi method \citep{1953ApJ...118..113C}
and follow Equation 2 in  \citet{2004ApJ...600..279C}:
\begin{gather}
    B_{pos}  {\chth =Q} \sqrt{4 \pi \rho} \frac{\delta V}{\delta \phi} \approx 9.3 \frac{\sqrt{n\left({H}_{2}\right)} \Delta V}{\delta \phi} \mu G,
\end{gather}
where $n\left({H}_{2}\right)$ is the molecular hydrogen number density of the region, $\delta V$  is the (average) velocity dispersion of the gas, $\sigma$ is the dispersion in the polarization position angles, {\chth we assume Q to be approximately 0.5 (see below)}, and $\Delta V = \sqrt{8~\rm{ln}(2)} \delta V$. To estimate the average density of the region we use our  C$^{18}$O data, as emission from N$_2$D$^+$ (a much higher density tracer) does not cover the entire region where polarization was detected (see \autoref{fig8}). 
{\chth We estimate the column density of C$^{18}$O by following equations in \citet{1991ApJ...374..540G} and \citet{2010MNRAS.401..204B}:
\begin{gather}
N_{\mathrm{C}^{18} \mathrm{O}}= \frac{{\chft 1.26} \times 10^{14}}{1-\exp \left(-{\chft 10.54} / T_{\mathrm{ex}}\right)} \frac{T_{\mathrm{ex}}+0.88}{J\left(T_{\mathrm{ex}}\right)-J\left(T_{\mathrm{bg}}\right)} \int T_{\mathrm{mb}} d v,\\
J(T)=\frac{h \nu}{k} \frac{1}{\exp (h \nu / k T)-1}
\end{gather} where $T_{mb}$ is the main beam brightness temperature, $T_{ex}$ is the excitation temperature and $T_{bg}$ is the background brightness temperature which is due to cosmic background radiation.} 
We adopt a C$^{18}$O to H$_2$ ratio of $1.7\times 10^{-7}$ \citep{2019ApJ...873...16H} and assume the depth of the region to be about {\chft 0.1}\,pc (which is about {\chft half} the width of the JCMT polarization map) to estimate an average density of the region of $n\left({H}_{2}\right) \sim {\chft 1.7} \times 10^4$ cm$^{-3}$. We made Gaussian fits to the  C$^{18}$O spectra in the region and obtained an average velocity dispersion ($\delta V$) of 0.6\,km\,s$^{-1}$. With a measured polarization angle dispersion ($\delta \phi$) of 36.6$^{\circ}$, and using Eq.~5, {\chth we find the magnetic field strength in the plane of the sky in the medium-density region traced by C$^{18}$O is about {\chft $50\,\mu$G.}} 

{\chth  \citet{2001ApJ...546..980O} conducted MHD simulations and found the value of $Q$ in Eq.~5 ranges between $0.46-0.51$ when the measured dispersion in the polarization angle is less than $\sim$25$^{\circ}$. For such low dispersion, the expected  uncertainty in the value of $Q$ is less than 30\,\% \citep{2004ApJ...600..279C}.  However, it is also important to note that different simulations  result in slightly different $Q$ values. For example, the results by   \citep{2001ApJ...559.1005P} and \citep {2001ApJ...561..800H}  give Q ranges between $0.3-0.4$. Moreover, the Chandrasekhar-Fermi method is not optimal for dense structures where gravitational forces dominate over MHD turbulence. Given these caveats, 
{\chft our} method should give us a rough estimate of the magnetic field strength 
with an uncertainty of factor of a few \citep[e.g.,][]{2005ASPC..343..166H}.  Our estimate is enough to provide a general idea of how the magnetic fields affect the gas dynamics in the region. }


{\chth Theory predicts that magnetic field strength scales with gas density, such that 
$B \propto \rho^\kappa$, where $\kappa$ may be as low as 0 and as high as 2/3 depending on the evolutionary stage and geometry of the (collapsing) dense structure \citep{2012ARA&A..50...29C}. Thus, the  magnetic field in the central filament, where the average density is $1.6 \times 10^6$ cm$^{-3}$, is expected to be about {\chft 50 to 970} $\mu G$ (i.e.,  higher than the magnetic strength in the lower density region traced by the C$^{18}$O observations).
Here we will assume that the B-field strength in the dense filament is about {\chft 510} $\mu G$ (a value between the two extremes estimated above).  
This value of the magnetic field strength is similar to that measured in structures in star forming regions with a density similar to our filament \citep[e.g.,][]{2017ApJ...838..121C,2020arXiv200704923G,2020arXiv201013503A,2020NatAs.tmp..159P,2020arXiv201101555W}). Using this value for the B-field strength, we then estimate the Alfv\'en speed ($V_A = B/\sqrt{4\pi \rho}$) to be about {\chft 0.54} km\,s$^{-1}$ in the dense filament. In this region, where the average velocity dispersion ($\sigma_v$) is 0.23 km\,s$^{-1}$, the magnetic mach number ($M_A = \sigma_v/V_A$) is approximately 0.4.  This implies that the magnetic field can have a slightly larger impact on the gas dynamics of the filament than the turbulence. 
 We also find that in this filament $\beta_B=2(M_A/M_s)^2\sim 0.4$, where M$_s$ is the sonic mach number which has a value of  1.0. This indicates, that as expected, the magnetic pressure in this region is greater than the thermal pressure   \citep[e.g.,][]{1996ApJ...466..814G,1999ApJ...520..706C,2012ARA&A..50...29C,2017ApJ...838..121C}. }

Another way to {\chth assess the role} of magnetic fields in the dynamical evolution of a region is to compare the gravitational energy to the magnetic field energy, given by \begin{gather}
E_B = \frac{1}{2}MV_{A}^2,
\end{gather}
where $V_A$ is the Alfv\'en speed and $M$ is the mass of the region. 
{\chth For the filament in our study,  this results in {\chft $E_B =3.4 \times 10^{43}$}\,erg, using the values derived above.
Therefore the gravitational energy is about {\chft two} times }  larger than the magnetic field energy (see \autoref{table:1}), indicating the filament is magnetic super-critical (i.e., the magnetic field strength in this region is not enough to support the filament against gravitational collapse). {\chth  Even though the magnetic energy is about an order of magnitude larger than the rotational energy of the filament, and as discussed above, the B-field in this region is likely to have significant influence in the gas dynamics of this region,  
 magnetic braking  should be negligible in LBS 23 as the magnetic field orientation is approximately perpendicular to the rotation axis of the filament.}

\begin{figure*}[tbh]
\centering
\makebox[\textwidth]{\includegraphics[width=\textwidth]{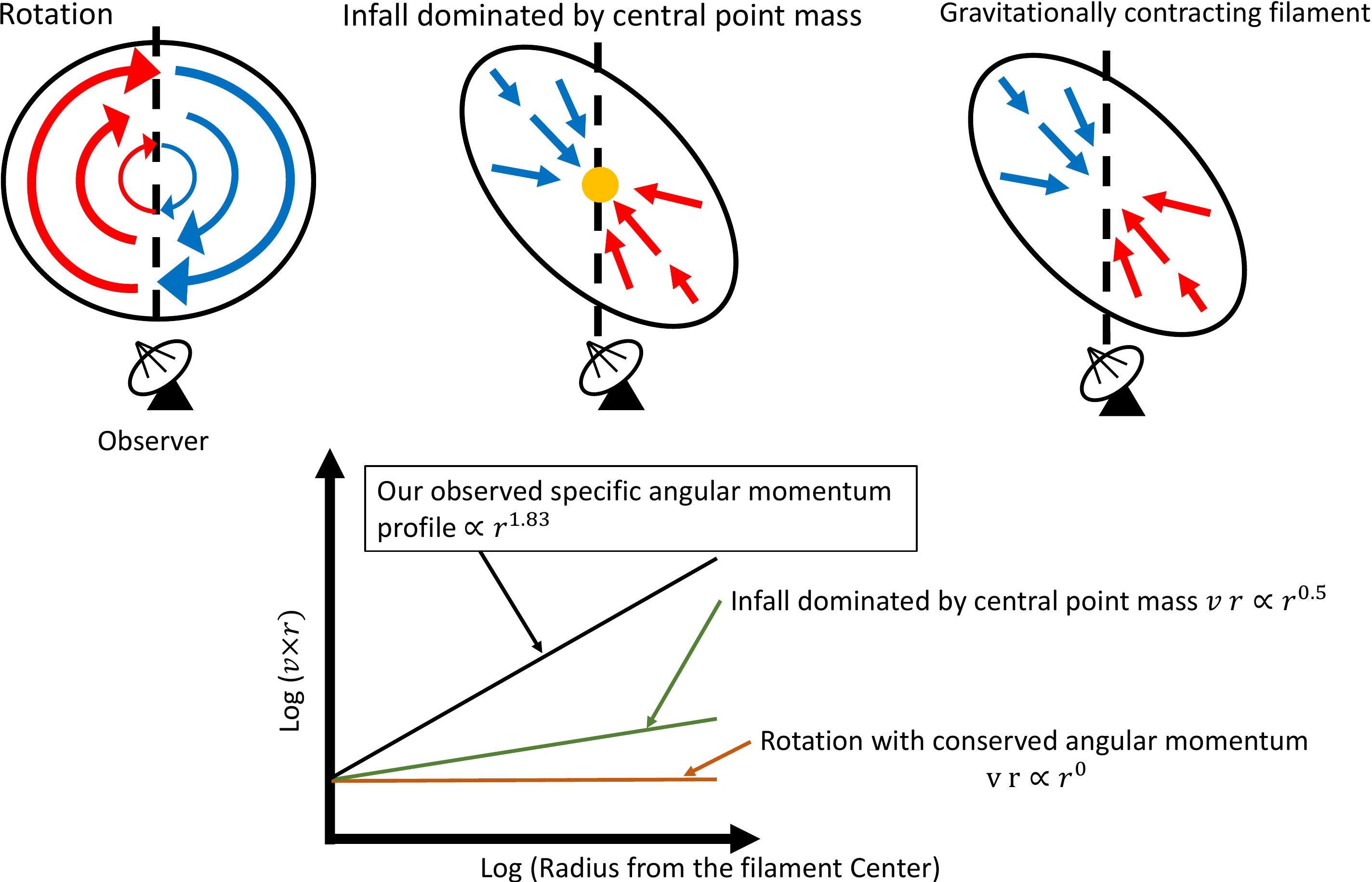}}
\caption{{\chtw 
Schematic representation
 of  collapse scenarios discussed in  Sec.~4.6: 
1) rotation with conserved angular momentum; 2) infall dominated by point mass; and 3) gravitationally contracting filament.
Diagrams are based on Figure 15 of \citep{2018ApJ...853..169D}. 
The bottom plot shows  a comparison of the expected  $r \times v$  profile for the first two models and the profile observed in the filament in this study.}} 
\label{fig9}
\end{figure*}

\subsection{{\chtw Other interpretations of velocity gradients in filaments  and the case for rotation}}
\label{sec:why_rotation}

{\chtw Different studies have interpreted a velocity gradient perpendicular to a filament's major axis  as being caused by different processes, including filament rotation \citep{2002A&A...392.1053O}, colliding flows  \citep{2013MNRAS.428.3425H}, multiple velocities components along  the line of sight \citep{2015A&A...584A..67B,2018ApJ...853..169D} and gravitational infall of gas onto filaments with an elliptical cross section \citep[i.e., infall with a preferred direction as opposed to isotropic infall,][]{2018ApJ...853..169D,2020MNRAS.494.3675C}. Clearly, care must be taken to reveal the true nature of the velocity gradient across a filament.}

{\chtw Changes in the mean velocity of the gas that appear as velocity gradients  along the short axis of filaments in low-mass star-forming regions have been recently reported by  various studies
\citep{2014ApJ...790L..19F,2018ApJ...853..169D,2020MNRAS.494.3675C}. An explanation for the existence of the observed velocity structure in several of these filaments is the existence of overlapping  multiple velocity components along the line of sight. 
A good example of this is the filament in the Serpens Main-S region  \citep[see Figure 11 of][]{2018ApJ...853..169D}, where it can be seen
that the emission at different velocities (i.e., different velocity components) lie next to each other, but also cross over at an angle. This is different to the filament in our study, as in the channel maps presented in \autoref{fig4} we do not see any sudden change in direction in the emission structure;  the main structure shifts slightly in position in consecutive channels but maintains a north-south direction (see Sec.~\ref{sec:results}). It would seem extremely unlikely that many sub-filaments with almost the same morphology and the same length would perfectly align next to each other with such a well-ordered velocity structure. We thus can  safely rule out that multiple velocity components is the origin of the velocity gradient seen in our filament.}

{\chtw Another explanation discussed in the literature for the observed velocity gradient along the minor axis of a filament is converging flows (e.g., due to compression in a turbulent medium, or compression triggered by external processes such as supernova explosions or stellar winds).  
In the converging flows scenario both low-density tracers that probe the environment outside the filament and high-density tracers that probe the filament 
itself are expected to show similar kinematic structures \citep[e.g.][]{2015A&A...584A..67B,2020MNRAS.494.3675C}{\chft .}
We inspected the intensity weighted velocity 
(moment 1) maps of our lower-density gas tracers in our ALMA observations (i.e., C$^{18}$O,$^{13}$CO, H$_2$CO, and $^{12}$CO) and we do not see any evidence of large scale flows towards the central filament. There is no clear velocity gradient perpendicular to the major axis of the central filament in these lower-density gas maps. 

In addition to this qualitative comparison,
we follow the prescription given by 
\citet{2020MNRAS.494.3675C} to determine whether the observed velocity gradient is  due 
to turbulent compression. 
If the dimensionless quantity given by 
\begin{gather}
C_v \equiv \frac{\Delta v_h^2}{GM/L},
\end{gather}
is significantly  greater than 1, then the velocity gradient in the filament is likely  
due to shock compression \citep{2020MNRAS.494.3675C}. In this equation
 $\Delta v_h$ is half the observed velocity difference across the filament minor axis, and $M/L$ is the filament's linear mass. In our case, we find the velocity difference between the east and west ends of the filament is $\sim 0.6\,$km\,s$^{-1}$ (see \autoref{fig5}), which   gives $\Delta v_h \sim 0.3$ km\,s$^{-1}$. 
 We estimate the linear mass of our filament within the region from which we obtain the velocity gradient (see \autoref{fig5}a)  
 to be 69.5\,M$_\odot$/pc, which results in  $C_v\approx 0.3$. We, therefore can rule out the scenario where the observed velocity gradient in LBS\,23 is 
 due to the convergence of large-scale flows or sheet-like structures created by turbulence compression.
 Even though, according to \citet{2020MNRAS.494.3675C} our estimate of $C_v$ should indicate 
 that self-gravity is important in shaping the velocity profile in the filament{\chth,} we argue below that although gravity is important in our filament, rotation is a more likely explanation for the observed velocity structure.
 

{\chtw A velocity gradient across a filament can also be produced by anisotropic infall in a filament formed inside a flattened structure or slab (see Figure 15 in \cite{2018ApJ...853..169D}; and Figure 1 in \cite{2020MNRAS.494.3675C}). 
In order to determine 
whether the velocity gradient in our filament is due to infall, we consider the expected velocity profile from three different simple (collapse) models that have been used to describe the velocity structure of cores and filaments: 1) a scenario where the  observed velocities are dominated by rotational velocities in a dynamically collapsing structure where angular momentum is conserved
\citep[e.g.,][]{1997ApJ...488..317O}; 
2) a model of radial  free-fall collapse (with no rotation) under the influence of a central point mass
\citep[e.g.,][]{1998ApJ...504..314M}
; and 3) a gravitationally contracting filament \citep[as discussed by][]{2020MNRAS.494.3675C}. In the first case, where angular momentum ($l=mvr$) is conserved, we expect the velocity profile to be $v\propto r^{-1}$. In the case of 
 free-fall collapse under a central point mass, radial velocities are governed by the conservation of energy  ($\frac{1}{2}mv^2=\frac{GM}{r}$), and we expect $v\propto r^{-0.5}$. 
 The infall velocity for a 
 gravitational contracting filament with a mass 
 as function of radius ($M(r)$), and a length $L$ 
 is given by $v^2 \approx G\,M(r)/L$ (see Equation 3 of \citet{2020MNRAS.494.3675C}). In \autoref{fig6} we fit the density profile and show that $\rho(r) \propto r^{-2}$. 
 In \autoref{sec:appendixB} we use the density profile of our filament to derive 
$M(r)= \pi L \rho_c R_{flat}^{2}\left[\ln \left(\frac{R_{flat}^{2}+r^{2}}{R_{flat}^{2}}\right)\right]$, which would result in an infall velocity profile of $v \propto \sqrt{\ln(1+(r/R_{flat})^2)}$ for our filament.

We compared the derived specific angular momentum profile of our filament with the specific angular momentum profile one would naively expect to detect if one were to assume that an observed velocity gradient in the three models described above were due to rotation
($j_{obs}(r)$). To do this we simply multiply the expected velocity profile of the model by $r$. Thus, for example $j_{obs}(r)$ for the second model described above would be  $vr\propto r^{0.5}$.
In \autoref{fig9} {\chth we} show schematic diagrams of $j_{obs}(r)$ for the first two models listed above and the specific angular momentum profile derived for the central filament in LBS\,23.
The slope (i.e., the power-law index) of the observed profile in our filament is significantly higher than these two collapse scenarios. 
In \autoref{fig7} we also compare the derived specific angular momentum profile for our filament and the
$j_{obs}(r)$ one would expect for 
a gravitationally contracting filament with a mass distribution similar to that of  the central filament in LBS\,23. Again we see that the derived 
$j_{obs}(r)$ for our filament is significantly different from that expected from the  model. It is therefore unlikely that the detected velocity gradient in the central filament is mainly due to gravitational infall. 

\citet{2016MNRAS.455.3640S} used  hydrodynamic turbulent cloud simulations to study the formation and kinematics of filaments in molecular clouds. From these simulations they were able to decompose the kinematic structure of the filaments into different components, one of which was the rotational velocity. Even though Smith et al. conclude that filaments that form in their simulations do not have ordered rotation on scales of 0.1 pc, they do detect rotational velocities of up to about 0.23 km s$^{-1}$ (similar to the maximum rotational velocity we detect in our filament of 0.3 km s$^{-1}$), and their filaments show ordered rotation at the scales of a few 0.01pc similar to the scales of our filament
{\chth \citep[see Figure~9 in][]{2016MNRAS.455.3640S}}. We are thus confident that rotation in a small filament like the one we studied here is possible. 

The velocity structure of the N$_2$D$^+$ emission allows us to confidently assert that the observed velocity gradient in the central LBS23 filament is not due to 
parallel sub-filaments at slightly different velocities.  Similarly, we are convinced that the filament velocity gradient is not caused by colliding flows since we do not detect 
velocity gradients across the filament at larger scales with lower-density gas tracers. Moreover, we discard gravitational collapse as the main cause of the observed velocity gradient as the derived specific angular momentum profile for our filament
 significantly deviates from that expected from three different collapse scenarios. We thus conclude that rotation is the most likely scenario as our filament's $j_{obs}(r)$  is consistent with the specific angular momentum profile observed for cores with velocity gradients that are generally presumed to be due to rotation.

\section{Summary And Conclusions}
\label{sec:conclusion}

We have analyzed the kinematic structure of a star-forming filament in the HH 24-26 region (a.k.a. LBS23) in Orion B, using ALMA N$_2$D$^+$ observations. The data clearly
shows a gradient along the filament's minor axis  which we argue is caused by rotation in the filament. 
From this we obtain a reliable estimate of the specific angular momentum in a rotating star-forming filament,  comparable to the specific angular momentum of cores with similar size found in other  star-forming regions. 

We compared the data with both rotating and non-rotating cylinder models, and found that in both cases the observed linear mass is higher than the critical linear mass above which the filament (cylinder) is expected to be unstable against collapse. 
Multiple dust continuum point sources at the ends of the filament, coincident with high-velocity outflow emission, suggest that there is ongoing star formation taking place in this filament, consistent with the measured high linear mass.

The dependence of the filament specific angular momentum profile as a function of radius ($j(r) \propto r^{1.8}$) is consistent with that observed in cores in other regions of star formation and provides evidence which suggests that the process that produces the velocity gradients in cores may be similar to what took place in this filament. 
The power-law dependence of the specific angular momentum with radius in the filament studied here is seen to continue down to scales of several $10^2$ au, with no indication of flattening. That is, there {\chth is} no detectable scale at which the angular momentum is conserved.  This suggests that 
the rotation in this filament may have been set by the turbulence in the cloud at all scales, even  down to the scale of the triple system that has formed in this filament.  This is consistent with  the scenario in which filaments and cores develop simultaneously from  the multiscale growth of nonlinear perturbations generated by turbulence, 
and it is in contrast with the traditional two-step scenario where thermally supercritical filaments form first and then fragment longitudinally into cores. 

Filaments, in general, fill the missing scales between cores and cloud. Thus, more measurements of filament angular momentum are needed in order to have a clear picture of 
how angular momentum is  transferred from cloud scales to cores.

We analyzed the turbulence in the central filament and found it to be transonic. This is in contrast with the expected subsonic motions in coherent cores and filaments, where turbulence decays on 0.1\,pc scales
The existence of a very young protostellar triple system in the filament suggests that star formation can occur even before turbulence decays down to subsonic motions. We further estimated the turbulence energy dissipation rate and found it to be at least an order of magnitude smaller than the typical outflow energy injection rate from protostars in a similar nearby cloud.  The excess of outflow energy injection rate may be able to sustain the turbulence in the filament and prevent from further collapse in the future.

Using archival  $850\,\mu$\,m JCMT SCUBA polarization data, we find the magnetic field on the plane of the sky in the LBS\,23 region is mostly perpendicular to the filament. 
{\chft The orientation of the magnetic field with respect to the filament's rotation axis  implies that  magnetic breaking effects should be negligible. 
 Using the Chanrasekhar-Fermi method we estimate the plane of sky magnetic field strength in the extended (medium-density) region surrounding the filament to be  
approximately $50\,\mu$\,G. Following theoretical predictions
which indicate that magnetic field strength increases with gas density, we speculate the magnetic field strength in the dense filament should be about a factor of ten larger ($\sim 500\,\mu$\,G), similar to the magnetic field strength determined in other similar high-density regions.}

\begin{acknowledgements}
      CHH and HGA acknowledge support from the National Science Foundation award AST-1714710. DM acknowledges support from CONICYT project Basal AFB170002. The authors thank Simone Recchi and Alvaro Hacar for conducting the calculation and providing the critical linear mass table for additional truncation radii. {\chel The authors further thank the NAASC Data Analysts, Sarah Wood and Tom Booth who helped with data reduction during the face-to-face visit.} This paper makes use of the following ALMA data: ADS/JAO.ALMA \#2016.1.01338.S ALMA is a partnership of ESO (representing its member states), NSF (USA) and NINS (Japan), together with NRC (Canada) and NSC and ASIAA (Taiwan) and KASI (Republic of Korea), in cooperation with the Republic of Chile. The Joint ALMA Observatory is operated by ESO, AUI/NRAO and NAOJ. The National Radio Astronomy Observatory is a facility of the National Science Foundation operated under cooperative agreement by Associated Universities, Inc. 
\end{acknowledgements}

{\cht \software{CASA (McMullin et al. 2007), astropy (The Astropy Collaboration 2013, 2018).}}

\begin{appendix}{}
\section{Column Density estimation}
\label{sec:appendix}

We use the Herschel-Plank dust optical depth map from \citet{2014A&A...566A..45L} and our \N2Dp ALMA map to estimate the H$_2$ column density of the filament, based on the method used by
\citet{2018A&A...610A..77H} using  N$_2$H$^+$ data and Herschel-Plank dust continuum emission maps in Orion A. The 850\,$\mu$m optical depth map, which has an angular resolution of  36\as0 over the region of interest, was first converted to K band extinction ($A_K$) using the expression:
\begin{gather}
A_K = \gamma \tau_{850}+\delta,
\end{gather}
where $\gamma_{Orion B}=3460$\,mag and $\delta_{Orion B}=-0.001$\,mag  \citep[see Eq.~11 in][]{2014A&A...566A..45L}.  Following \citet{2017A&A...606A.123H},
we then find the column density by converting the K-band extinction to V-band extinction, 
using $ A_V/A_K = 8.933$ \citep{1985ApJ...288..618R}, and converting $A_V$ into H$_2$ column density, using $N({\rm H}_2)/A_V = 0.93 \times 10^{21}$ cm$^{-2}$ mag$^{-1}$ \citep{1978ApJ...224..132B}. 



We then compare our \N2Dp data with the derived column density map in order to estimate the mass of  the  filaments. To do this, we first produced an integrated intensity map of the  ALMA \N2Dp Total Power data by integrating 
over velocities where there is significant emission (i.e., from $V_{lsr} = 8.4$\,km\,s$^{-1}$ to about  11.3\,km\,s$^{-1}$). We then smoothed the \N2Dp integrated intensity map,
which has a resolution of 28\as2, to match the resolution of the derived column density map, using the CASA command \textit{imsmooth}. 
We  regrided both maps so that they have the same (equatorial) coordinate and (Nyquist-sampled) pixel scale, and 
obtained the  values of the column density and the  \N2Dp integrated  intensity (W(${\rm N_2D^+}$)) for each position (i.e., pixel). In \autoref{fig10} we show a scatter plot of these values, where a clear correlation between the H$_2$ column density and  \N2Dp integrated intensity is detected. A line fit to the data:
\begin{gather}
N({\rm H}_2) = m\times W(({\rm N_2D^+}) + C
\end{gather}
gives the best-fit parameters as  $m = (8.97\pm 0.56) \times 10^{24}$ cm$^{-2}$\,(Jy\,arcsec$^{-2}$\,km\,s$^{-1}$)$^{-1}$ and $C =(1.62 \pm 0.15) \times 10^{22}$\,cm$^{-2}$. {\chtw 
We then applied this empirical relationship to the high resolution \N2Dp integrated intensity (momnet 0) map (\autoref{fig1}a). The final (high-resolution) column density map is shown in \autoref{fig1}a. The mass can then be obtained using this map and the following formula:
\begin{gather}
M({\rm H_2})  = 2 \mu m_p N({\rm H_2}) A ,
\end{gather}
where $\mu=1.37$, $m_p=1.67\times10^{-24}$ g is the proton mass, and $A$ is the area of interest.



\begin{figure*}[tbh]
\centering
\makebox[\textwidth]{\includegraphics[width=\textwidth]{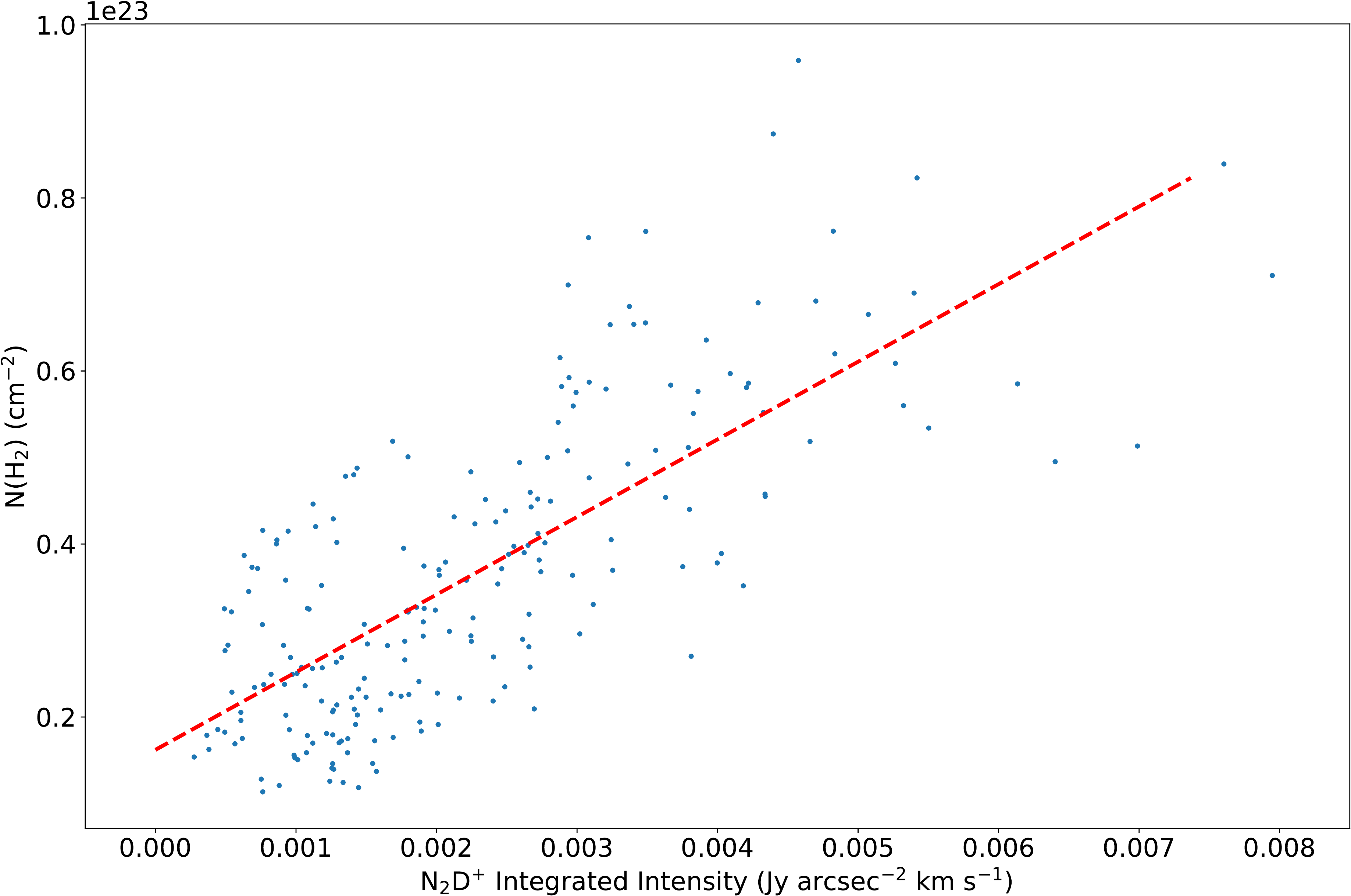}}
\caption{Empirical relation between the H$_2$ column density  map obtained from the Herschel-Planck dust continuum map and the N$_2$D$^+$ integrated intensity from the ALMA total power observations.
Each point represents the value of N(H$_2$) and W(N$_2$D$^+$) at the same position in the maps. The red dash line shows the line fit to the points. 
}
\label{fig10}
\end{figure*}

\section{Abundance ratio estimation}

We follow the formalism in \citet{2002ApJ...565..344C} to estimate the \N2Dp column density  from the line emission. For an optically thin line, the column density can be expressed as:  
\begin{gather}
N_{N_{2}D^{+}}= \frac{C}{J_\nu(T_{ex})-J_\nu(T_{bg})}\frac{1}{1-e^{h\nu/kT_{ex}}} \frac{Q_{rot}}{g_le^{-E_l/kT_{ex}}}
\end{gather}

where \begin{gather}
C =\frac{8\pi W}{\lambda^{3}A_{ul}}\times \frac{g_l}{g_u}\\
J(T)=\frac{h\nu}{k}\frac{1}{exp(h\nu/kT)-1}\\
Q_{rot}= \sum_{J=0}^{\infty} (2J+1)exp(-E_J/kT)\\
E_J=J(J+1)hB
\end{gather} 
In the equations above $Q_{rot}$ is the partition function\footnote{Note that the partition function of rotational transitions neglecting the hyperfine structure is different from the hyperfine partition function  \citep[see Appendix A in][]{2019A&A...629A..15R}. We use the former as
we do not resolve the individual hyperfine lines
in our observations 
of the N$_2$D$^+$ $J=3-2$ rotational transition.},   W is the integrated intensity of the line (in K km s$^{-1}$), $B=38554.719$\,MHz is the rotational constant \citep{2002ApJ...565..344C}, and {\chel  $A_{ul}=\,7.138\times10^{-4}$\,s$^{-1}$ is the Einstein coefficient for the $J=3-2$ transition \citep{2009A&A...494..719P,2019A&A...629A..15R}.} In the calculation, we assume the excitation temperature is 10\,K. After obtaining the {\chel column density map of \N2Dp}, we divide the total column density estimated from the Hershal-Plank map (see Appendix A above) to find the \N2Dp/H$_2$ abundance ratio, which we show  in \autoref{fig11}b. {\chtw The \N2Dp/H$_2$ abundance ratio ranges between $(6-8)\times10^{-12}$, which is similar to the 
\N2Dp abundance observed in other dense cold regions  \citep[e.g.,][]{2020ApJ...895..119T}. }

\begin{figure}[h]
\centering
\includegraphics[width=\hsize]{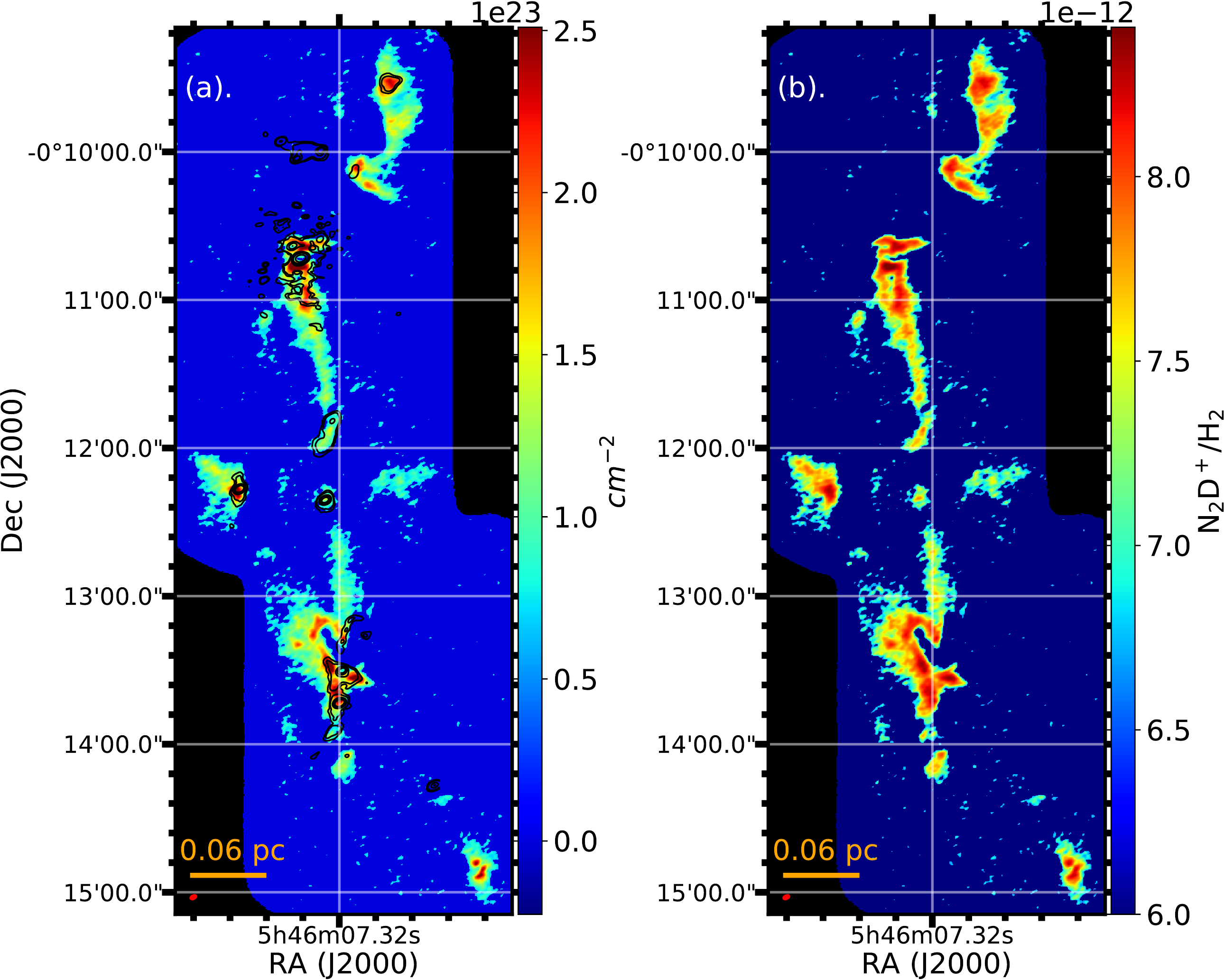}
\caption{(a).
High-resolution H$_2$ column density map obtained using the ALMA high-resolution \N2Dp integrated intensity  map (\autoref{fig1}a) and 
the empirical relation derived in
\autoref{fig10}.
 Black contours show the 1.29\,mm dust continuum emission in steps of 3$\sigma$, 5$\sigma$, 20$\sigma$, 40$\sigma$, 80$\sigma$, 320$\sigma$, where $\sigma = 5.4\times 10^{-4}$ Jy beam$^{-1}$ 
  (b). Map of the \N2Dp/H$_2$ abundance ratio. 
  The size of the synthesized beam is represented with a red ellipse in the lower left corner of each panel.}

\label{fig11}
\end{figure}

\newpage

\section{ Gravitational and Rotational Energy Estimation }
\label{sec:appendixB}
Here we describe our procedure for estimating the rotational and gravitational energies  of the filament.  
 Consider a rotating cylindrical filament with mass $M$, radius $R$ and length $L$. The rotational axis is along the direction of the cylinder's length. The rotational energy is given by:}
{\chtw \begin{gather}
    E_{rot}=\frac{1}{2}I\omega^2
\end{gather}
}



{\chtw  In our case, we need to obtain
the moment of inertia ($I$) for a cylinder with a non-uniform density.  The equation for
the surface density of
an idealized cylindrical filament with a Plummer-like profile is:
\begin{gather}
    \Sigma(r)=   \frac{A_p \rho_c R_{flat}}{\Bigg[(1+(\frac{r}{R_{flat}})^{2})\Bigg]^{(p-1)/2}},
\end{gather}
and the corresponding radial density profile is given by:
\begin{gather}
\rho_{\mathrm{p}}(r)=\frac{\rho_{\mathrm{c}}}{\left[1+\left(r / R_{\mathrm{flat}}\right)^{2}\right]^{p / 2}},
\end{gather}
where $$
A_{\mathrm{p}}=\frac{1}{\cos i} \int_{-\infty}^{\infty} \frac{\mathrm{d} u}{\left(1+u^{2}\right)^{p / 2}}
$$ \citep{2011A&A...529L...6A}. We consider our filament to be on the plane of sky ($i= 0\arcdeg$). 
In Sec.~4.3 we fit 
the column density of the central filament with a Plummer-like profile given by Eq.~3 (see \autoref{fig6}). 
Using $p=2$ (see Sec.~4.4), we find $A_p=\pi$. Using our estimate 
of $N_0$  and $R_{flat}= 0.006$\,pc (both obtained from our fit to  Eq.~3), and using $\Sigma = \mu m_H N_{H_2}$, where $m_H$ is the hydrogen mass and $\mu = 2.33$ is 
the mean molecular mass \citep{2011A&A...529L...6A},
we then obtain an estimate for the central density, $\rho_c$ (in Eq.~C11),  of
$9.88\times10^{-18}$\,g\,cm$^{-3}$.

We  then use the  density profile to obtain the momentum of inertia ($I$):
\begin{gather}
I = \int_{0}^{R} r^2 \rho(r) dV = \int_{0}^{R} r^2 2\pi r \rho(r) L d{r}\\
= 2 \pi L \int_{0}^{R} r^3 \times \frac{\rho_c}{1+(r/R_{flat})^2}  d{r}\\
= \pi L \rho_{c} R_{flat}^{2}\left[R^{2}-R_{flat}^{2} \ln \left(\frac{R_{flat}^{2}+R^{2}}{R_{flat}^{2}}\right)\right]
\end{gather}

The mass of a cylindrical filament with a density profile given by Eq.~C11 can be obtained with the following equation:
\begin{gather}
M = \int_{0}^{R}  2\pi r \rho(r) L d{r}\\
= 2 \pi L \int_{0}^{R} r \times \frac{\rho_c}{1+(r/R_{flat})^2}  d{r}\\
=\pi L \rho_c R_{flat}^{2}\left[\ln \left(\frac{R_{flat}^{2}+R^{2}}{R_{flat}^{2}}\right)\right]
\end{gather}
Using Eq.~C17 in Eq.~C14 we can get an expression for $I$ in terms of the filament mass $M$ (and independent of $L$):
\begin{gather}
I=\frac{M}{\ln \left(\frac{R_{flat}^{2}+R^{2}}{R_{flat}^{2}}\right)}\left[R^{2}-R_{flat}^{2} \ln \left(\frac{R_{flat}^{2}+R^{2}}{R_{flat}^{2}}\right)\right].
\end{gather}

For our filament  $M = 11.7$ M$_\odot$, $R_{flat}=0.06$\,pc,  the average radius is $R=2770$\,au and $\omega=6.5\times 10^{-13}$\,rad\,s$^{-1}$. Hence, the rotational energy of the filament is $3.0\times 10^{42}$\,erg.}


{\chtw The gravitational energy in a filament is given by:
\begin{gather}
E_G = -\frac{GM^2}{L},\\
\end{gather}
where, again, $M$ and $L$ are the mass and length of the filament \citep{2000MNRAS.311...85F}.
For our filament $L\sim 34,000$\,AU, which results in an estimate of the gravitational energy of  $-7.1\times 10^{43}$\,erg. Therefore, the ratio of the rotational energy to gravitational energy ($\beta_{rot}$) is $\sim 0.04$.


}

\end{appendix}

\end{document}